# Title: Principles of Physiological Closed-Loop Controllers in Neuromodulation

**Authors: Victoria S. Marks[1], Joram van Rheede[2], Dean Karantonis[3], Rosana Esteller[4], David Dinsmoor[5], John Fleming[2], Barrett Larson[6], Lane Desborough[7], Peter Single[3], Robert Raike[5], Pierre-François D'Haese[8, 9], Dario J. Englot[10], Scott Lempka[1,11], Richard North[12], Lawrence Poree[13], Marom Bikson*[14], Tim J. Denison*[1,2]**

***These authors contributed equally**

**Affiliations:**
1. Institute of Biomedical Engineering, Department of Engineering Science, University of Oxford, Oxford, UK; 2. Medical Research Council Brain Network Dynamics Unit, University of Oxford, Oxford, UK; 3. Saluda Medical, Macquarie Park, NSW, Australia; 4. Department of Biomedical Engineering, University of Minnesota, Twin Cities, MN (starting Aug 25); 5. Medtronic, Inc., Minneapolis, MN, USA; 6. Department of Anesthesiology, Perioperative and Pain Medicine, Stanford University School of Medicine, Stanford, CA; 7. Nudge BG Inc, Los Angeles, CA, USA; 8. Rockefeller Neuroscience Institute, West Virginia University, Morgantown, WV, USA; 9. Upstream Vision, New York City, NY, USA; 10. Department of Neurological Surgery, Vanderbilt University Medical Center, Nashville, TN, USA; 11. University of Michigan, Ann Arbor, MI, USA; 12. Neurosurgery, Anesthesiology and Critical Care Medicine, Johns Hopkins University School of Medicine (retired), Baltimore, MD, USA; 13. University of California San Francisco, San Francisco, CA, USA; 14. Department of Biomedical Engineering, The City College of New York, New York City, NY, USA

**Abstract:**
As neurostimulation devices increasingly incorporate closed-loop functionality, the greater design complexity brings additional requirements for risk management and special considerations to optimise benefit. This manuscript creates a common framework upon which all current and planned neuromodulation-based physiological closed-loop controllers (PCLCs) can be mapped including integration of the "Technical Considerations of Medical Devices with Physiologic Closed-Loop Control Technology" guidance published in 2023 by the United States Food and Drug Administration (FDA), a classification of feedback (reactive) and feedforward (predictive) biomarkers, and control systems theory. We explain risk management in the context of this framework and illustrate its applications for three exemplary technologies. This manuscript serves as guidance to the emerging field of PCLCs in neuromodulation, mitigating risk through standardized nomenclature and a systematic outline for rigorous device development, testing, and implementation.

**Objective:**
This document provides a tutorial on physiological closed-loop controller implementation within neuromodulation and provides a checklist aligned with FDA guidelines.

**Learning Objectives:**
1. Provide a unified framework with clear terminology of neuromodulation including physiologic closed-loop controllers
2. Explain and apply concepts from FDA guidance documents and PCLC standards (e.g. risk, 60601-1-10) for robust design and use
3. Develop a system to analyse "closed-loop" (PCLC) systems with a clear mental model of their operation
4. Use examples from multiple sensing-based domains to provide intuition for PCLC operation under this framework

## Table of Contents

## Introduction

Our approach to defining key terms used in the control of neuromodulation devices is to follow the 2023 FDA "Technical Considerations for Medical Devices with Physiologic Closed-Loop Control Technology" which references the IEC 60601-1-10 requirements for medical electrical equipment. The definitions used here should be understood as further limited only to the context of applications in neuromodulation control. We also strove to maintain continuity between definitions in the glossary by [1] and risk management paradigms outlined in [2], but we have expanded upon both to include specific applications. This tutorial aims to provide a useful resource for stakeholders across the device development pipeline, from physicians, scientists, and engineers to patients and the general public, to understand how best to apply PCLCs to their indication such that it works predictably, successfully mitigates risk, and improves therapy efficacy.

A *Physiological Closed-Loop Controller (PCLC)*, as defined by IEC60601-1-10, is a medical device or system that automatically adjusts or maintains a physiologic variable(s) through delivery or removal of energy (e.g., electric) or matter (e.g., drugs, or liquid or gas considered as a medical device) using feedback from a physiologic-measuring sensor.

PCLCs rely on proper integration of biomarkers, or objective, measurable physiological variables that meaningfully reflect a biological process, disease process, or response to a therapeutic intervention. The evolution of PCLCs in neuromodulation is positioned at the convergence of stimulation and sensing technologies and increasingly sophisticated control algorithms, including leveraging artificial intelligence (AI), representing a significant shift toward truly personalised and adaptive therapies. Recent advances in control algorithms have enabled the development of more sensitive, accurate, and reliable biomarkers – both reactive and predictive – that enhance real-time decision-making capabilities of neuromodulation systems. For instance, the integration of machine learning methods for feature extraction and classification has substantially improved the detection and predictive accuracy of critical physiological states, such as seizure onsets or pain exacerbations, facilitating timely and precise intervention adjustments.

A critical frontier in the field is the expansion of closed-loop neuromodulation from primarily reactive systems towards fully autonomous adaptive therapies leveraging predictive AI. Emerging capabilities such as multi-modal sensor integration, real-time edge computing, and deep learning are enabling closed-loop systems not only to respond to immediate physiological signals but also to anticipate clinical needs based on patterns discerned from longitudinal data. Innovations such as evoked compound action potential (ECAP)-adjusted spinal cord stimulation (SCS) illustrate how neural biomarkers can directly drive dynamic adjustments in stimulation dose to maintain therapeutic efficacy despite physiological variability, significantly improving patient outcomes and device usability[3]. Moreover, adaptive deep brain stimulation systems that modulate therapy based on beta band power in Parkinson's disease (PD) demonstrate the clinical viability of major loop PCLCs to continuously adapt to disease-specific neurophysiological states.

### Stimulation Dose

Stimulation *dose* is defined as those aspects of technology that impact how energy is applied to the body[4]:

a) *Amplitude* of the stimulation, which is the peak intensity, whether voltage-controlled (expressed in volts) or current-controlled (expressed in amperes)

b) *Waveform and timing* of the stimulation, which spans how long therapy is applied, the interval between treatments, and how intensity changes during treatment (such as *frequency* and *pulse duration*).
c) *Location* where energy enters the body. Location refers here only to the relevant interface between the device and body (not all device parts). For electrical stimulation, location is the size, shape, and placement of electrodes with respect to the target neural tissue. Changing location can therefore change the energy requirements in (a) and (b).

Neuromodulation dose and the properties of the body together determine which cells are exposed to what energy, which in turn drives the outcome of neuromodulation. In this sense, two devices that provide an identical dose to the target neural tissue are indistinguishable to the body – they are the same in the therapeutic responses they produce. Device parameters outside of dose, such as its user interface and battery life certainly matter but are not part of the therapeutic dose. Therefore, when we personalise neuromodulation dose, we adjust only amplitude, location, or waveform/timing.

Similar to pharmacokinetic dose responses, the *electrophysiokinetics*, or the relationship between the amplitude and waveform of the electrical stimulation and the response of the target tissue, may be non-linear or non-monotonic. For example, stimulation that depends on neural activation must exceed the activation threshold to have benefit and stimulation below that value will not be efficacious. Stimulation may affect the body via multiple mechanisms, including longer term neuroplasticity processes, which can have beneficial or counterproductive components. The dose/response relationship for each of these mechanisms should be established to determine how to provide predominantly beneficial, and preferably optimal, therapy.

Each neuromodulation device is designed to provide a limited range (set) of neuromodulation doses. Each neuromodulation device limits its range of dose parameters based on technology (e.g., hardware limitations), the use-case (intended use), and safety standards[5-7] (e.g., limiting charge density below that which could cause tissue damage). This does not mean it is advisable to try any dose available from a given device; for example, some doses may be painful. A single neuromodulation device may allow a few or many possible doses – that is, combinations of amplitude, location, and waveform/timing. Given too many dose possibilities to practically test and a cost to testing ineffective doses (e.g., lost time, side effects), there must be a therapy technique or algorithm to search for and select a preferred neuromodulation dose regimen.

Individualised therapy with most drugs is limited to coarse adjustment of amount (e.g., number of pills) or timing (e.g., morning and evening). Although, adaptive pharmaceutical devices do exist, such as the artificial pancreas, a combination of a constant-glucose monitor and an insulin pump, which allows blood-sugar control using feedback[8]. Neuromodulation devices are inherently designed to allow dynamic, real-time dose adjustment. This allows therapy to be personalised, even on a treatment-by-treatment basis. Dose can be adjusted mid-treatment because there is an instant corresponding change in energy in the body. The pharmacokinetics of drug delivery are sluggish by comparison. However, it is important to note the exact time relationship between energy delivery and symptom reduction varies by indication, and some (e.g., depression, epilepsy[9]) may require weeks or months to see an effect from neuromodulation.

By adjusting stimulation location, neuromodulation can deliver energy to one or several neural targets, in contrast to systematically distributed drugs. Changing stimulation location allows engaging distinct regions of the nervous system that are implicated in disease aetiology or its control. A device location may be changed by moving electrodes, for example adjusting a transcutaneous electrical nerve stimulation device over a painful body region. In situations where a device has several electrodes already placed, stimulation location may be changed by selecting which contacts are activated – which is the case for implanted devices.

Thus, neuromodulation devices are adjustable in real-time and with specificity in both time and region. When dose is adjusted for an individual, it is personalised. The selection of one out of the many possible doses relies on biomarkers measured from individuals. The next two sections describe how biomarkers are used in personalised neuromodulation dose tuning, including classifying distinct approaches and classes of biomarkers.

**Manual versus Automatic Dose Adjustment**

The traditional strategy for neuromodulation is to set certain fixed stimulation parameters, monitor the patient's symptoms, and manually change the settings during clinical visits as needed. This "manual loop" is considered *human **in** the loop*: all changes to therapy are implemented directly by a clinician, nurse, or, to a limited extent, the patient. There is no automated "control policy" embedded on the device. Stimulation parameters are programmed based on clinician or patient input, and changes in behaviour are reported by a patient through self-report or in a clinical exam. If available, complementary monitoring sensors (e.g., electrophysiology) indicate if any changes to the physiology result. Changes to either physiology or behaviour are fed back to and interpreted by the *human **in** the loop* to update or keep fixed the stimulation dose. This method often sets chronic stimulation parameters (for use over weeks/months) on single snapshots in time (e.g. determined on a clinical examination or programming day) but has limited ability to adjust stimulation dose on an ongoing and acute basis. In addition to the clinician, the *human **in** the* loop can include the patient, who is provided with a programmer with limited control on dose (e.g., on/off intensity) as pre-set by the clinician.

Newer devices, with automated loops, will detect, predict, and react to state changes in the physiological system as shown in Figure 1. This "automated loop" still involves a *human **on** the loop* who monitors its operation: a clinician sets the initial conditions of the loop (set point, dose, stimulation parameter limits, and basic control rules), but the device has the capacity to make changes to the stimulation dose while the loop is running without necessary intervention from the clinical team. In this case, changes in the physiological variable are directly compared to a target, or *setpoint*, and the difference, or *error,* is fed back to the algorithmic control to react and change stimulation parameters. Devices may measure and adapt to physiological data (biomarkers) or non-physiological data (e.g., time of day, battery level). Devices may respond to the current state of the measured variable or a predicted future state. Among biomarkers, we explain below the distinction between those used in feedback and feed-forward control. Automated loops also include *fallback modes*, which begin in response to unexpected values from any of the feedback or feedforward signals, and emergency stops. The *human **on** the loop* can also include the patient with important but restricted programming capabilities.

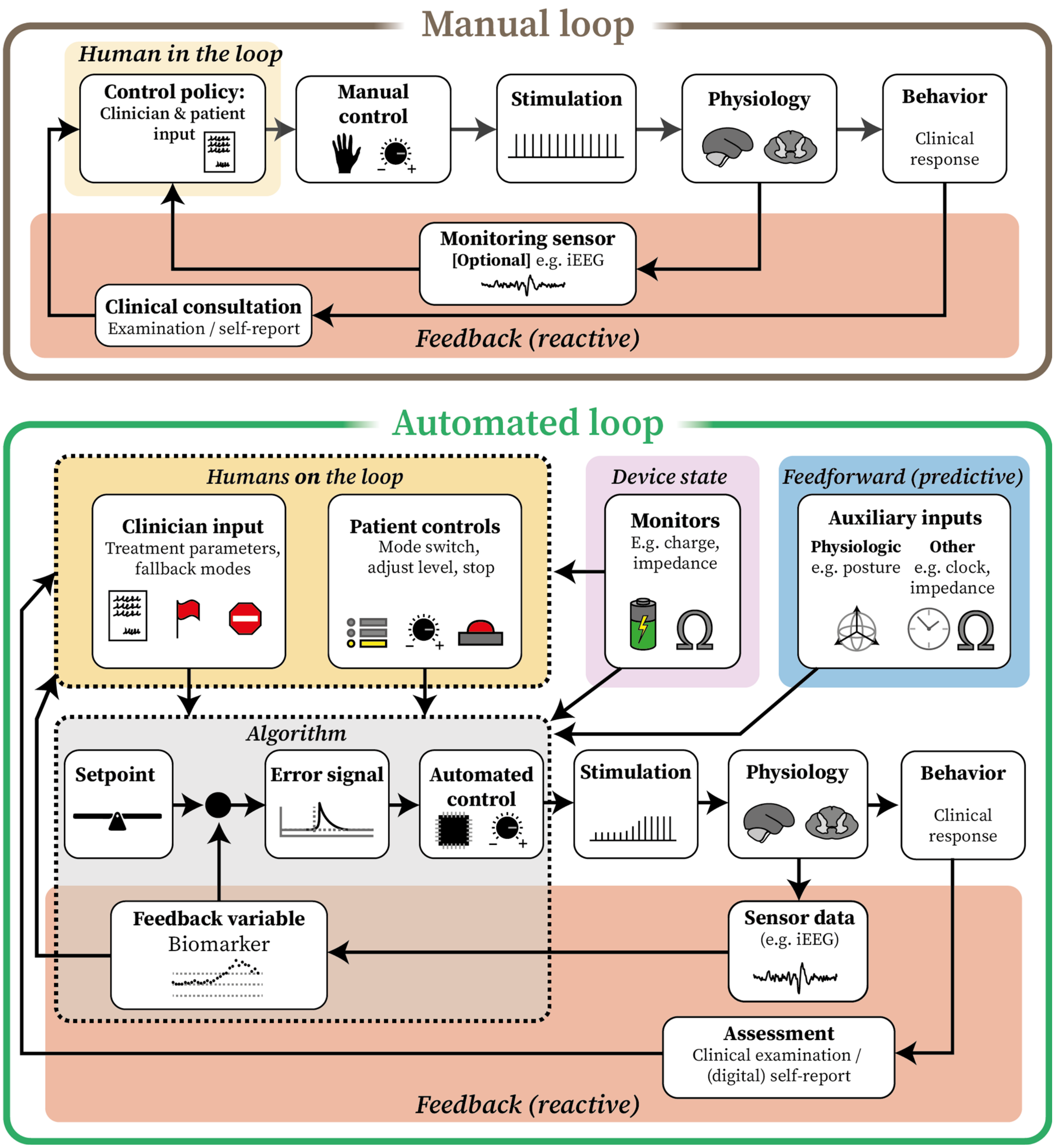

**Figure 1.** Physiologic closed-loop controllers can be either manual or automated loops depending on the level of human involvement: whether the actuator is fully controlled by the human in the loop (manual loop) or if the actuator runs automatically within human input guidelines and periodic monitoring of device logs and alerts to ensure proper running (automatic loop). Clinicians input all the rules for the algorithm box, including stimulation limits, flags, and fallback modes. Patients may be able to switch programs manually, fine tune amplitudes, or activate emergency stops to stimulation. Automated PCLCs allow for incorporation of more flexible algorithms and more complex variable integration (i.e., both feedback and feedforward elements) while also monitoring device state (e.g. battery life, electrode integrity).

Moving from a manual to an automated loop does not reduce the level of responsibility of human members of the clinical team, but it has the potential to reduce their frequency of interaction. While in both manual and automated loops, the clinician is responsible for configuring sensors, selecting biomarkers, and monitoring for alerts, the crux of the difference between the two approaches is how therapy is changed over time. In a manual loop, the role of the clinical team involves selecting and troubleshooting the exact dose and stimulation paradigm, including stimulation location, time, frequency, pulse width, and waveform (usually a variation of a rectangular pulse) to optimise therapeutic dose. As disease symptoms fluctuate with time, the provider and patient select a predetermined follow-up frequency to adjust these stimulation settings as needed. In an automated loop, however, the clinical team sets the initial stimulation paradigm, and this paradigm then adjusts stimulation dose in an automated fashion over time. The provider, however, is still required to set limits on the stimulation parameters, configure algorithm control policy parameters, and meet regularly with the patient to ensure optimal delivery of therapy. Human involvement would also be required if an entirely new stimulation paradigm were desired, or if an automatic mode were stopped for any reason. Thus, the use of an automated loop does not diminish the role of the clinical team nor lessen the imperativeness that they maintain a good mental model of the automated processes at work. IEC 62366 defines mental models as "how users think the system behaves vs how it actually behaves," and IEC 60601-1-10 Section 4.2.2 requires the designer to "ensure that the user interface communicates the control system's current state, mode, and intent in a way that matches the operator's mental model." For this reason, each of the practical PCLC examples below includes a schematic of the basics of the mental model for that application, building on the general framework in Figure 1.

**Types of Biomarkers**

Automatic control relies on feedback (*reactive*) biomarkers and/or feedforward (*predictive*) biomarkers. In this section, biomarker types are explained based on a prior framework and analysis[10]. Biomarkers, as considered here, are specifically used to adjust stimulation dose and are part of control loops for dose optimisation. The nature of the biomarker impacts how it is used in a loop. Through the inclusion of both feedback and feedforward variables, we expand beyond the IEC 60601-1-10 definition of a PCLC "command variable" as purely feedback against a "comparing element."

*Reactive biomarkers* (also called responsive biomarkers) are measures of physiological activity that are expected to respond directly to stimulation and can therefore inform adjustments to the stimulation paradigm. For example, if the reactive biomarker is in the desired range compared to a *setpoint*, the stimulation dose would maintain a constant level. While, if the reactive biomarker is not in the desired range compared to a setpoint, the stimulation dose would be changed according to a pre-specified control policy. Once a new dose is selected, changes in the reactive biomarker continue to be monitored. Reactive biomarkers can be broadly separated into three types. *Type 1* reactive biomarkers are clinical response surrogates: durable responses to dose which track changes to primary clinical outcomes and are sustained after stimulation is off. *Type 2* reactive biomarkers typically concern the mechanism of action; they measure an acute change in the body's response to dose. *Type 3* reactive biomarkers are instant, electrodynamic changes that measure energy (e.g., electric field) delivered to the tissue.

Whichever type of reactive biomarker is used, it is important to characterise the dose response curve (Figure 2), paying special awareness to what may cause differences from the ideal response (Figure 2A) in offset, gain (change in biomarker per change in stimulation amplitude; Figure 2B), noise, and tonicity (Figure 2C). Figure 2D-F show examples of dose response curves from SCS and from DBS for Parkinson's, highlighting how these factors affect the region of operation (usable stimulation amplitudes) for a given device and application. Our scope is to provide common definitions and explanations at the tutorial level to learn about the field; more nuanced, application-specific considerations are beyond the scope of this manuscript.

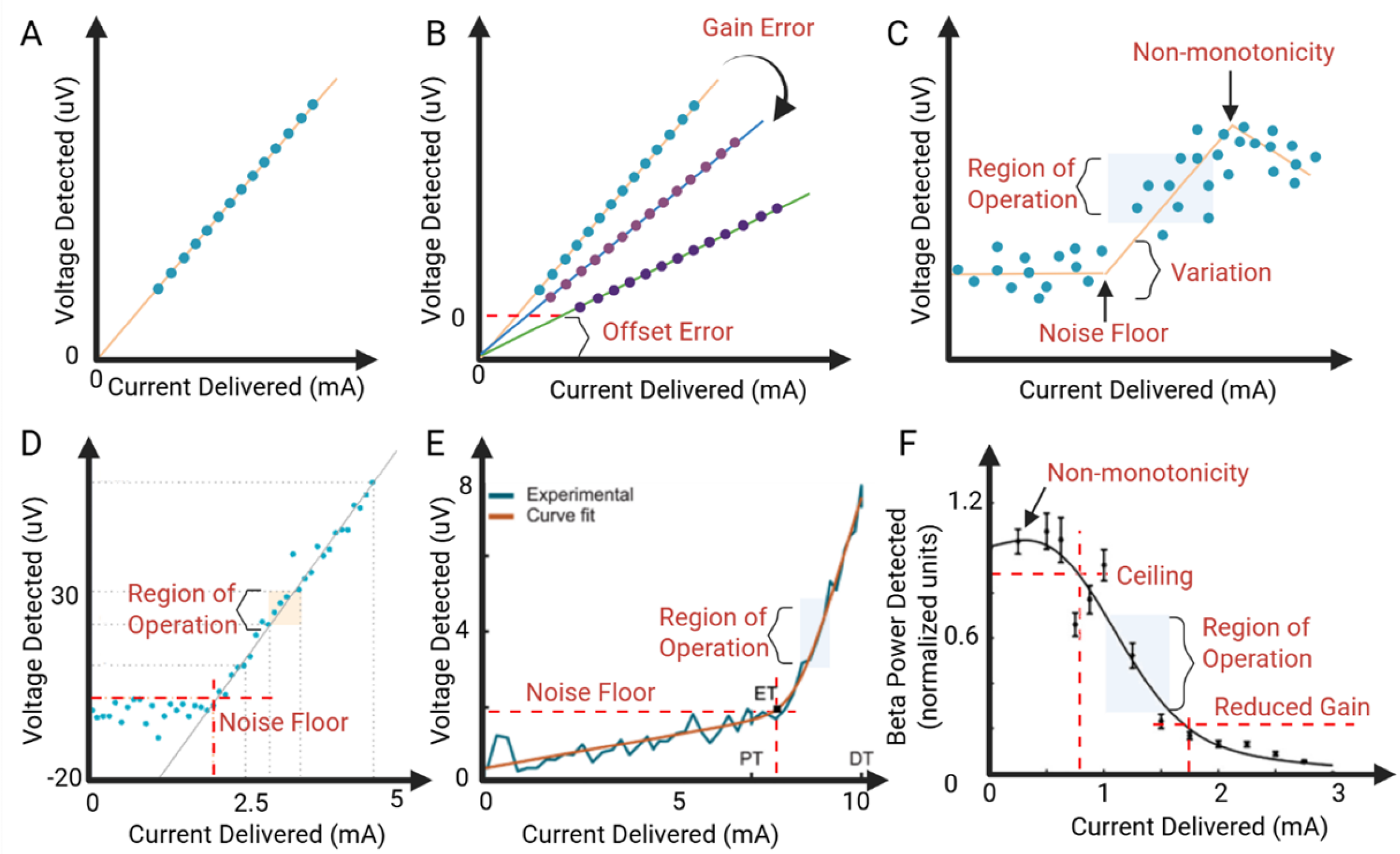

**Figure 2. Biomarker design considerations.** A. Ideal, noiseless, monotonic dose response. B. Schematic of incorporation of offset error and gain error. C. Schematic of incorporation of noise, variation, and non-monotonicity. D and E. Adaptations of data from Saluda (D) and Medtronic[11] (E), highlighting general similarities between different devices, although they may still have different variability and dose response. F. Adaptation of a figure modelling beta dose response[12], based on real data from Parkinson's patients[13], highlighting that some biomarkers may decrease with dose, and gain is not constant. In this case, there is diminishing biomarker reduction with increased amplitude around 1.75 mA.

*Predictive biomarkers* are measures of physiological activity or anatomy that are not expected to directly respond to stimulation. Predictive biomarkers can be measured before stimulation starts or during stimulation (e.g., genetics (DYT-1) can predict dystonia response)[14 15]. Stimulation dose is adjusted based on the predictive biomarker without the expectation the predictive biomarker will change. For example, stimulation intensity may be changed based on subject posture without the expectation that posture will be influenced by stimulation. Predictive biomarkers may be "gates" that determine the timing of dose delivery (e.g., syncing stimulation pulses to breathing). Alternatively, predictive biomarkers may be "playback" of a specific signal feature which provides information on whether stimulation dose and location are optimised. Predictive biomarkers may be considered "neuronavigation" if they provide information on the position of neuroanatomy in relation to dose delivery – in this way, fluoroscopy is considered a neuronavigation biomarker predictive of effectiveness of dose delivery. Evoked responses to test stimulation can also be used as predictive biomarkers that may indicate change in brain excitability or brain state and would then provide information on how stimulation dose should or should not be altered. Predictive biomarkers may be measured once or repeatedly.

It is important to note these are simplified categories of biomarkers based on how they are used[10], such that the same biomarker may be used in one system as reactive (feedback) and in another as predictive (feedforward). Some biomarkers may have overlapping characteristics between types, and some PCLCs may actuate based on multiple biomarker inputs. There could be any number of combinations of acute, durable, reactive, and predictive biomarkers monitored operating on any number of timescales (i.e., latency to response, settling time). For this reason, a hierarchy of biomarkers (i.e., an explicit definition of biomarker priority) and how a device responds to them is a critical design consideration.

**“Major” versus “Minor” Loop Control**

Neuromodulation PCLCs can operate with different hierarchies of feedback control for appropriate titration of parameters such as stimulation amplitude, pulse width, and frequency. Control systems engineers refer to these hierarchies as “major loop” control which operates around the entire system under control, and “minor loop” control, which closes the loop around a sub-component. Each control method can provide therapy improvements. An example of major loop control is to sense a signal directly associated with a disease state—such as elevated beta band power, a pathophysiological marker of PD—and adjust the therapy in response to changes in this signal[16] (see example below for more details). Likewise, a PCLC can leverage minor loop control, reacting to a signal not strictly associated with a disease state but still important to maintain the integrity of the system. For example, in SCS systems, minor loop feedback can help automatically optimise parameters by making the system more immune to variations in the tissue-electrode interface. Owing to post-operative maturation of the electrode-tissue interface and spinal cord motion with postural shifts, the number of activated fibres resulting from the stimulation can shift over time and during activities of daily living[17 18].

In its simplest form, constant-current stimulators that adapt the output voltage based on feedback from impedance measurements are employing minor loop control; in this case, the improved immunity to impedance variations can stabilise the charge delivery not as a PCLC but as an important subcomponent of the PCLC design. More advanced technological mitigations for number of activated fibres variability now include posture-responsive stimulation circuitry using evoked potential measurements and feedback control systems[3 19] (see example below for more details). While minor loop control does not use a symptom biomarker as the feedback control variable, positive therapeutic impact may nevertheless be realised through their incorporation, particularly in the context of system automation and therapy optimisation relevant to dose consistency. For long-term therapy roadmaps, researchers continue to explore direct pain biomarkers[20] which might one day enable a major loop control around a primary symptom of interest.

**Risk Mitigation**

A key takeaway from the FDA guidance and the 60601-1-10 framework is the critical need for proper risk mitigation at each step of the loop (as described in Figure 3), including variable logging. The first step is to have an intuitive mental model through which to understand the patient transfer element, or the relationship between the physiological signal sensed and the electrical stimulation delivered, especially in comparison to the steady-state value of the physiologic variable. This mental model includes identification of the type of reactive or predictive biomarker sensed, how different types of stimulation affect biomarker behaviour, and which levels of stimulation are safe or unsafe. This information helps set functional risk mitigation parameters such as physiological sensing limits and stimulation limits. The next step is to be mindful of the possibility that a device malfunctions or an unexpected artefact is encountered as an implanted person interacts with their environment. The FDA guidance warns against several pitfalls of PCLC implementation, specifically complacency, loss of situational awareness, automation bias, and skill degradation[21].

The likelihood of failure from either complacency or loss of situational awareness is reduced by implementation of proper monitoring (e.g., logs, alerts) and setting of entrance and exit criteria. It is important to consider every possible failure mode, to monitor data that could indicate such failures (e.g., impedance, device temperature), to alert when human intervention may be necessary, and to plan for the next appropriate action to take for each failure mode. In such cases that a failure mode is encountered that meets the loop's exit criteria (evidence such that it is unclear whether appropriate adaptive stimulation is being provided), an automated PCLC should switch to a fallback mode (e.g., a stimulation setting within the safe, therapeutic range which does not change with the biomarker). Keep in mind that a fallback mode may, in some cases, mean turning stimulation off or reverting to manual control. Whichever fallback mode is initiated, it should continue operating until the next time a log indicates that the failure mode has been resolved, and the loop's entrance criteria is met. In some cases, a fallback mode may be initiated by the patient or physician (e.g., by a magnet swipe) during a clinical visit (e.g., while undergoing imaging) or other instance in which a specific, sustained setting is required. While not all failure modes can be prevented, their likelihood of occurrence and potential impact on therapy delivery are reduced through the implementation of alerts, state logging, actuation limits, thorough consideration of dynamics, and fallback modes.

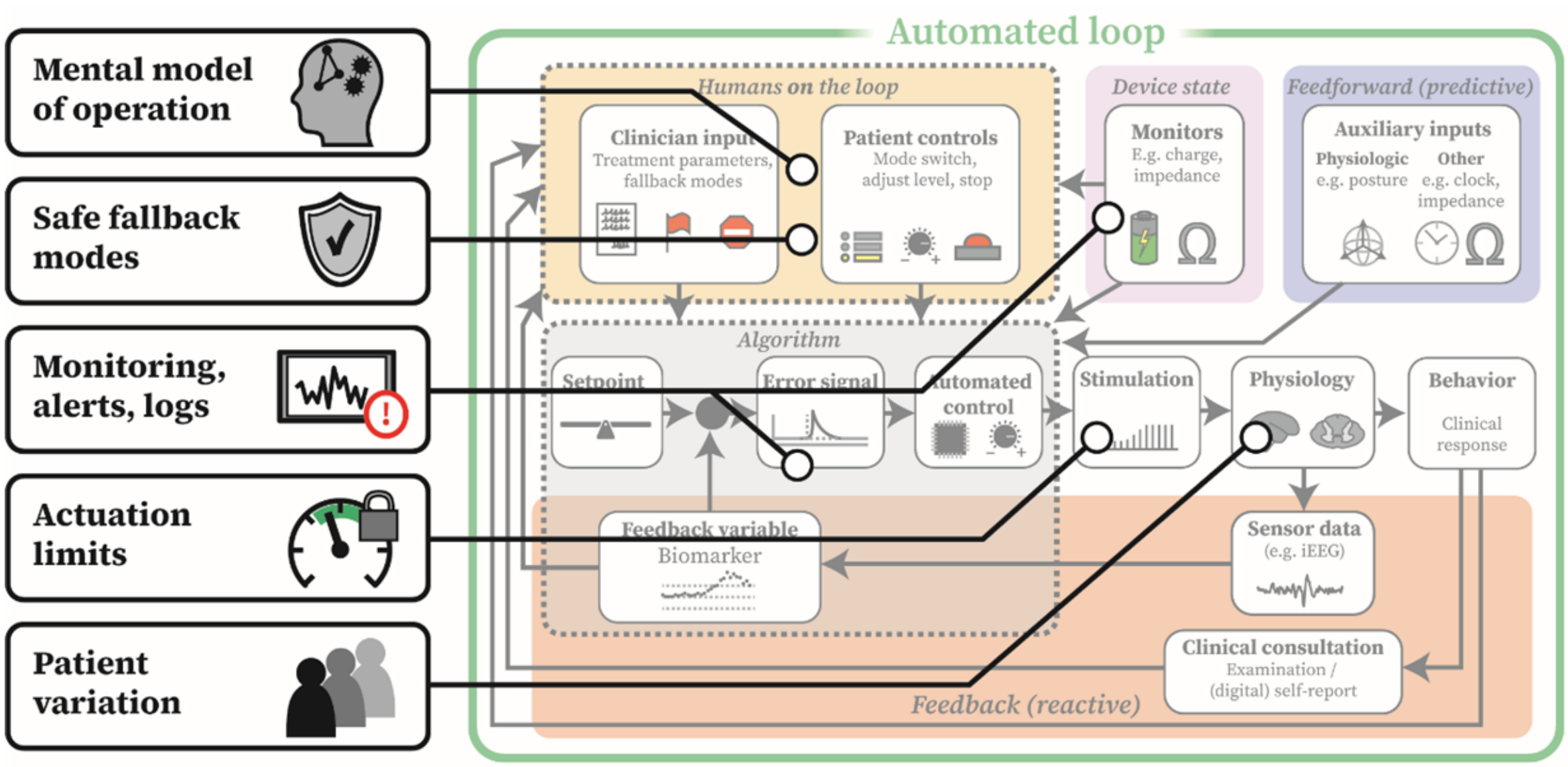

**Figure 3:** Risk must be mitigated at every step of PCLC implementation. In automated loops, the user sets actuation limits and safe fallback modes with appropriate monitoring, alerts, and logs to flag anytime the system violates the mental model of operation, conflicts with assumptions about physiological limits, or otherwise requires human intervention (e.g., battery charge).

The likelihood of failure from either automation bias or skill degradation is further reduced by proper verification and validation testing. Because human physiology is dynamic, all settings are subject to both inter- and intra-patient variation, including the value of the setpoint. System settings should be verified to accommodate this dynamic range and have proper monitoring in place to alert if something is outside of this expected range or requires human intervention. It is the clinician's job to check monitors and logs to make sure devices are still operating within the expected range. Before delivering devices for human use, safety measures should be rigorously validated from the bench, and then further validated in clinical trials before delivery for commercial use. Introducing automation into therapy delivery introduces inherent risks as the PCLC performs a task based on a sensed variable, without a human in the loop. Whilst the performance characteristics can and should be verified with *in silico, in vitro* and/or preclinical models, clinical evaluations should be conducted to assess the clinical value and risk profile of the technology. The clinician needs to understand how the device operates when it encounters a failure mode once deployed, not just when it is operating under ideal conditions.

Risk will never truly be reduced to zero, but implementation of good risk mitigation measures can get close. In the unfortunate case that an adverse event does occur, good monitoring and logging will aid in forming reports for the FDA and other regulatory bodies to review. These records are instrumental in identifying design deficiencies that lead to faults and allow for timely design improvements. Clinicians using these devices should be able to put these logs into the context of their mental model of the system and work to prevent repeating the adverse event in the future.

## Practical Examples

In the following sections, we discuss three practical examples of FDA-approved PCLCs used commercially. We explain the motivating principles and physiology, mental model, biomarkers for automated control, risk mitigation strategies, and practical implementation and performance metrics. These are simplified descriptions of the major components and loops of each use case, but keep in mind that any number of "minor" loops could be operating as well.

### Use Case: Responsive Neural Stimulation

This example introduces the NeuroPace RNS® System, the first PCLC device for responsive neurostimulation approved by FDA for neuromodulation to treat drug resistant epilepsy in 2013. The RNS® System delivers targeted neural stimulation in response to physiological changes detected in the intracranial electroencephalography (iEEG) or electrocorticography (ECoG) signals. Figure 4 adapts the general PCLC block diagram introduced in Figure 1 to this use case, illustrating how specific variables from the RNS® System map onto the generic framework.

*Motivation and Mental Model: Delivering Dose Responsive to a Biomarker for Epilepsy*

The mental model behind responsive neurostimulation therapy assumes that stimulation, when delivered near seizure onset, may interrupt or shorten seizure duration, reduce the likelihood of subsequent seizures, or both. Early evidence supporting this mental model came from Lesser et al.[22], who demonstrated that brief bursts of electrical stimulation could terminate after-discharges triggered by cortical stimulation in humans. The control loop described here has a bang-bang output, meaning one of two, discrete stimulation states: on or off. The RNS® System also supports two advanced responsive-stimulation paradigms: 1) frequency-adaptive stimulation, which automatically adjusts the burst frequency based on the detected iEEG signal, and 2) phase-synchronised stimulation, in which stimulation is delivered at a specified phase of the iEEG signal in the detection channel. Of note, adaptive stimulation with the RNS® System is still responsive therapy. It allows for automatic adjustment of the frequency of the responsive stimuli depending on the detected signal; this is differentiated from the continuous stimulation automated loop (also often called adaptive stimulation) for other systems and applications.

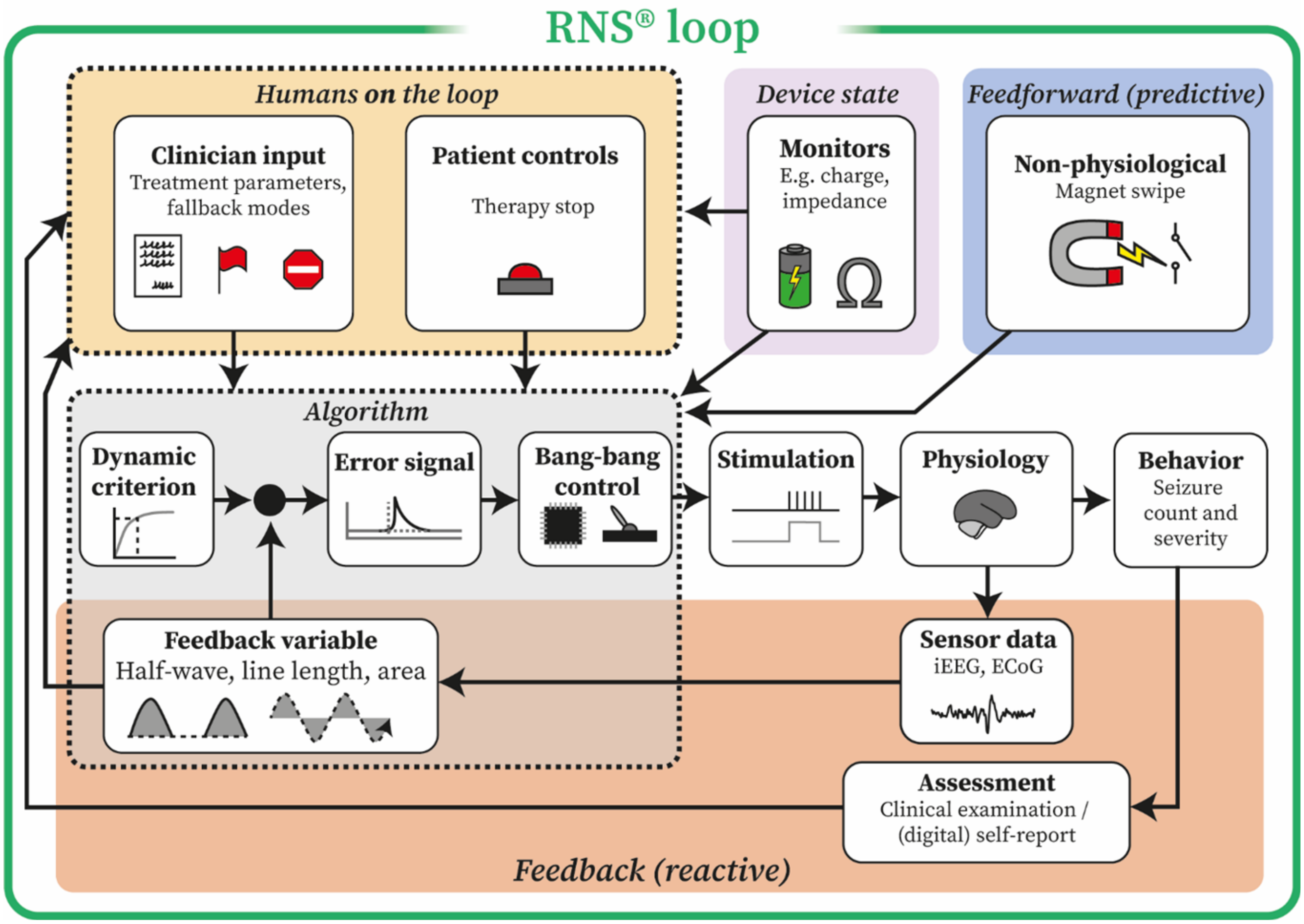

**Figure 4.** Block diagram of a general PCLC system, adapted to the specific examples of the variables used in this case study of responsive neural stimulation using the RNS® System for treatment of epilepsy. In this case, patient controls are limited to emergency stops or iEEG storage initiated by a magnet swipe. Instead of a stationary set-point, there are several dynamic criteria that influence the stimulation control. Monitors, including event logs, help inform future algorithm adjustments.

*Biomarkers for Automated Control*

The RNS® System comprises an implantable neurostimulator and leads, a patient remote monitor, a clinician programmer, and cloud-based data storage accessible 24/7 to authorised clinicians or researchers via any internet-connected smart device (Figure 5). The implanted device continuously analyses iEEG/ECoG signals and applies up to three seizure detection tools using patient-specific half-wave, line length, and area (area-under-the-curve) algorithms configured for early seizure detection[23 24]. These tools can be applied to one or two iEEG/ECoG channels individually or in combination with logical operators (e.g., OR, AND). When any selected feature exceeds its programmed threshold (fixed or variable threshold, Figure 4), it serves as a biomarker of epileptiform activity or an imminent seizure and triggers stimulation. We categorise these features as Type 1 reactive biomarkers as our mental model[22] relies on them as surrogate metrics of the main clinical outcome, seizure likelihood, however they share characteristics with Type 2 reactive biomarkers as the stimulation delivered results in a short-term change that may eventually become a durable change. The RNS® System uses a hybrid setpoint strategy: adaptive or fixed threshold for line length and area, and fixed threshold for half-wave. Adaptive thresholds are continuously updated based on a long-term baseline window compared to the window used to compute the short-term line length or area, which helps maintain sensitivity while minimizing false detections. Initial validation of these biomarkers was conducted using clinical data from the Emory University database, which was later integrated into the IEEE EEG portal database[25]. This validation formed part of the pre-submission materials required by the FDA for Investigational Device Exemption approval of the RNS® System feasibility study[26].

*Risk Mitigation Strategies for an Epilepsy PCLC*

The system architecture incorporates hierarchical layers of control (Figure 5). The embedded layer, housed in the implant, corresponds to the automatic loop in Figure 4 that manages detection and therapy delivery using a bang-bang control policy, effectively running the PCLC algorithm to ensure immediate responses to physiological changes. The two upper layers (monitoring and database layers) are manually operated and map to the “humans on the loop” block in Figure 4. Each layer communicates upward by providing data and downward by enforcing safeguards against risks and hazards, while enabling flexibility for future system enhancements. As data are gathered from multiple implantable neurostimulators across patients, AI-driven analyses at the database layer can be used to design monitoring algorithms that detect and warn of potential hazards, thereby reducing patient risk[27]. These algorithms may evolve into autonomous PCLC algorithms operating at the higher layers, typically with longer response times but relying on robust computing power that is not feasible at the implant level (embedded layer, Figure 5). Researchers can update the implantable device's closed-loop control policy at the database layer using all the data received from multiple devices to improve the PCLC algorithm and deploy it through the monitoring layer to the implantable device, effectively enhancing control over periods spanning hours, days, or months as the learnings and implementation updates are executed.

A key feature of the system is its safety-oriented design. As part of the embedded layer (Figure 5) or algorithm block (Figure 4), the system includes a manual safety override mode that allows patients to temporarily suspend therapy by placing a magnet within 1 inch of the implant, provided this capability is not disabled during programming[28 29]. Therapy automatically resumes once the magnet is removed.

For each event detected, the system enforces a safety limit by restricting stimulation to a maximum of five consecutive “therapies” per stimulation ”event” while the detection flag remains active[28 29]. In this case, each “therapy” consists of 1 or 2 bursts, and if 2 bursts are configured, they are delivered consecutively with no intervening delay.[29]. This controlled stimulation strategy is designed to minimise the risk of overstimulation. If the detection flag remains active after the maximum number of therapies has been delivered, stimulation is suspended and will not resume until the detection flag resets (typically indicating the end of the seizure event) and a new detection occurs[29]. In addition to a limit on the number of therapies per episode, the number of stimulation episodes per day is limited to a user-configurable maximum.

The RNS® System incorporates several automated fallback modes to ensure patient safety in the event of critical faults. These include the End-of-Service (EOS) reset, triggered when battery voltage falls below the safe threshold, causing the device to suspend therapy, detections, and all measurements. The general neurostimulator reset occurs if the system detects an unrecoverable internal error or reaches EOS, placing the device in a non-therapeutic state until reprogrammed or replaced. Additionally, a DC-leak reset is automatically initiated in response to electrosurgical interference, suspending all therapy and detections until clinical reprogramming is performed. Each of these states represents a fallback mode characterised by fault-driven, autonomous transitions to a predefined safe configuration[28 29].

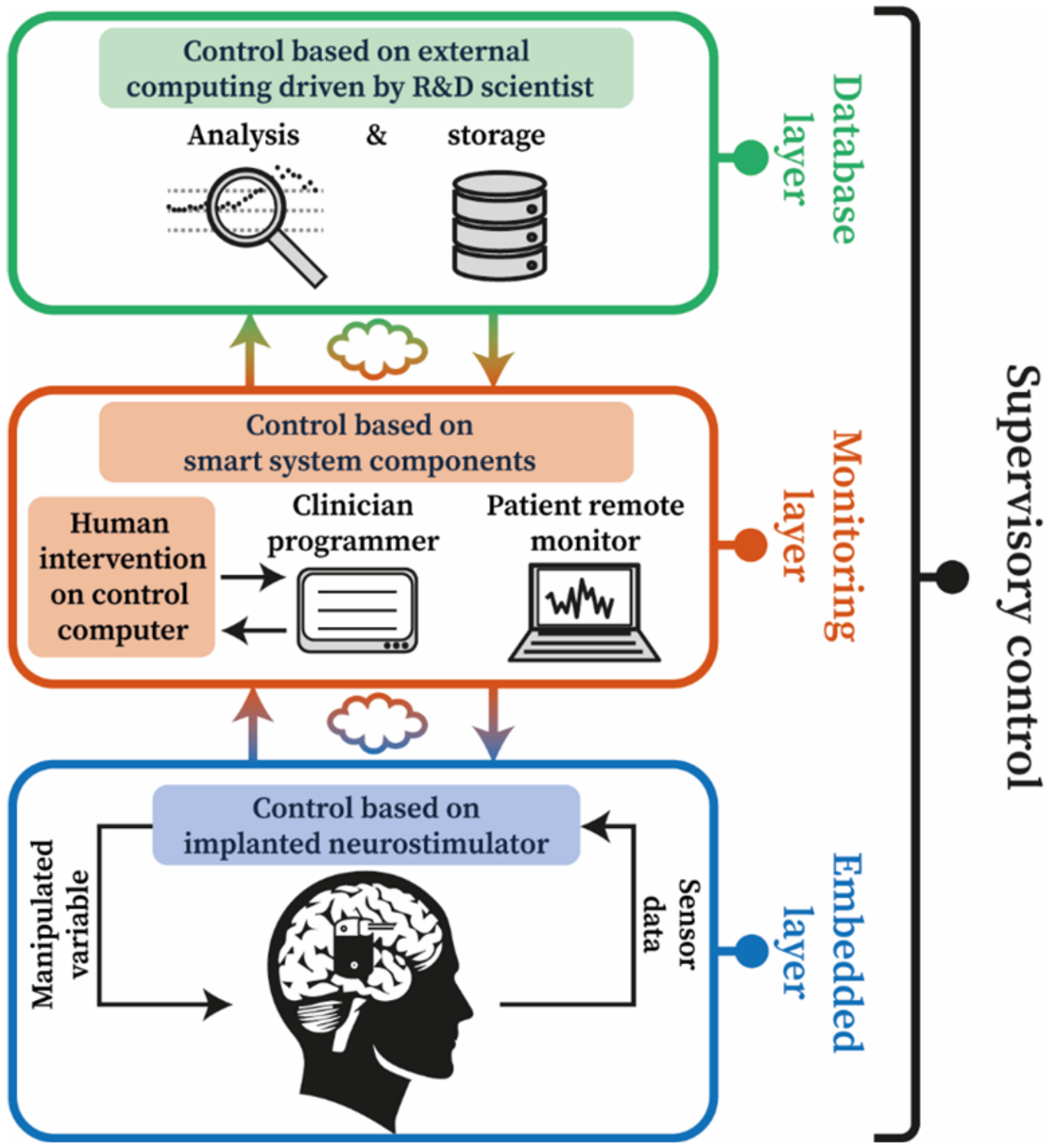

**Figure 5:** Example of robust and versatile PCLC system three-layer functionality (adapted from [27]).

*Practical Implementation and Performance*

The RNS® System was one of the first neuromodulation platforms to integrate a secure cloud-hosted Patient Data Management System (PDMS), enabling ongoing research and algorithm refinement[30-32] with the potential to update the system in the future, through a population-level database for optimisation[27]. The neurostimulator's onboard memory stores up to 32 minutes of iEEG data, with smart prioritization triggers to prevent overwriting critical recordings. It also stores 28 days of hourly event counts, reflecting clinical changes and cyclic patterns (circadian and multidien), and 1-2 days of time-stamped event records, providing a detailed record of episode times and durations. Patients and caregivers can transfer data to the remote monitor post-event to ensure physiological data retention. This stored data supports offline analysis and tuning of detection parameters. Users can simulate detection changes and visualise their effect on event classification through the PDMS platform, accessible via any connected device[28].

The system supports two depth or cortical strip leads (four contacts each), allowing targeted stimulation in one or both brain hemispheres. While it lacks independent current sources per contact which can lead to uneven charge distribution across variable impedances, it remains clinically effective. These design trade-offs reflect a balance between miniaturization, power constraints, and safety, all while delivering therapeutic performance across multiple seizure focus sites or network nodes. In other words, good feedback ensures stimulation adjustments until therapy works, making the operational performance of individual subcomponents less critical. Location of the implantable pulse generators (IPG) is also important when considering susceptibility to cardiac artifact. For example, the RNS® System is unique in its cranially mounted design makes it less susceptible to these than systems with pictorially mounted IPGs[33 34].

System performance has been tracked over time, with the most recent results reported in a prospective, multi-centre multi-year post-approval study[35]. Among the 324 patients included in this trial, 255 participants were included in the primary effectiveness endpoint: the overall seizure reduction was 82%, with neocortical onset patients achieving a 90% reduction and those with mesial temporal lobe epilepsy (MTLE) showing a 73.5% reduction.

*What's Next?*

The RNS® System sets a foundational precedent as the first closed-loop neuromodulation device demonstrating that physiological responsive therapy can be safely and effectively implemented in patients. Its combination of continuous physiological monitoring, cloud-integrated analytics, and AI-driven personalization has created a feedback-rich ecosystem for nurturing ongoing therapy improvements in the future[31 32].

Looking forward, as AI and automation capabilities grow, therapies like the RNS® System underscore the importance of modular architecture with data storage capabilities to support incremental innovation. The medical industry benefits from platforms that enable rapid iteration, such as the one in Figure 5 which allows components to evolve independently without requiring simultaneous development and is aligned with the PCLC framework we introduce here in Figure 1. Modular designs with relative component independence and architectural versatility enable greater integration flexibility, scalability, and adaptation to new clinical insights, market demands, and regulatory requirements. This strategy not only facilitates iterative advancement across device generations but also mitigates technical and business risks.

**Use Case: Parkinson's Adaptive Regulation**

*Motivation and Mental Model: Stabilizing Dose with Regulation of a Biomarker*

Deep brain stimulation (DBS) is a well-established therapeutic approach for alleviating the motor symptoms associated with PD[36 37]. Despite its efficacy, however, patients often continue to experience moment-to-moment symptom fluctuations. These fluctuations can be attributed to changes in physical activity, medication state (particularly with agents like levodopa), and circadian rhythms. Importantly, these variations are reflected in the neural local field potentials (LFPs) recorded from the brain, which show clear alignment with the patient's sleep-wake cycle[38] and contribute significantly to signal variability.

The ability to record neural signals directly from implanted stimulation electrodes presents an opportunity to apply a physiologic closed-loop controller (PCLC) to DBS systems. In accordance with the principles of IEC 60601-1-10, this kind of control system could provide real-time responsiveness to the patient's changing physiological state. Although patient-operated controllers are currently used to manually adjust stimulation amplitude, this method is limited by its practicality. Regular manual adjustment, particularly during nighttime or sleep, places a considerable burden on the patient and may fail to adequately respond to symptom variability. Furthermore, clinical adjustments made during routine visits are inherently limited by their infrequent and static nature, and some bursts occur on timescales (seconds) too fast upon which to respond. These limitations underscore the need for a more dynamic, automated approach to DBS delivery. As illustrated in Figure 6, the mental model of an automated loop for PD is to regulate the physiological biomarker analogous to a home thermostat.

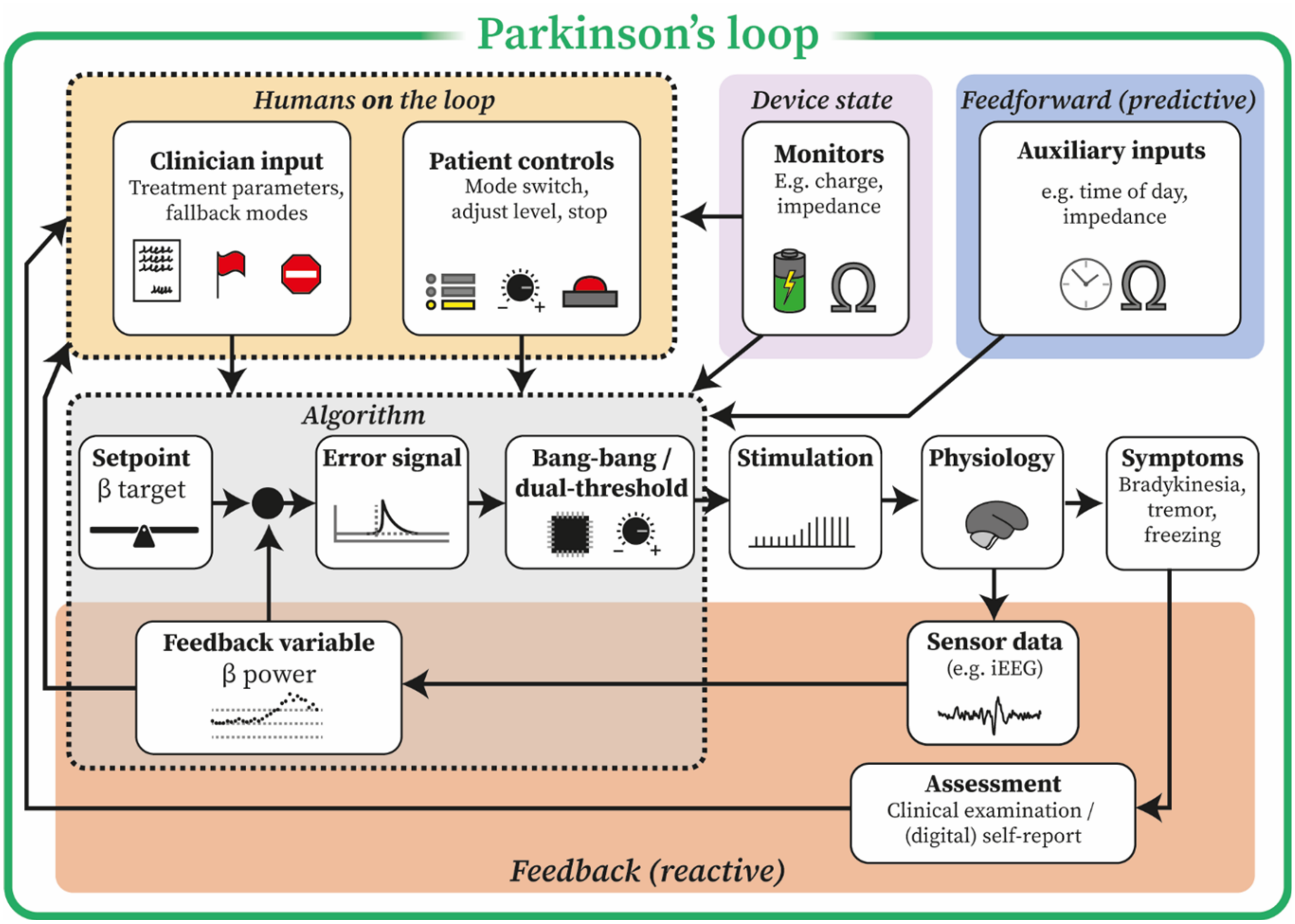

**Figure 6.** Block diagram of a general PCLC system, adapted to the specific examples of the variables used in this case study of deep brain stimulation for the treatment of Parkinson's Disease. Beta power is the classic reactive biomarker for a PD application, but other auxiliary inputs could be useful as predictive variables.

*Biomarkers for Automated Control*

LFPs are biopotential measurements of change in charge within the extracellular space, representing the summation of synaptic activity in the population of neurons around the recording contacts. LFPs recorded from the electrodes provide a window into the brain's ongoing activity, making them ideal candidates for use as biomarkers in adaptive systems. These signals, representing aggregate neural activity, can serve as both reactive (Type 2) and predictive biomarkers. Among the most common metrics used are the spectral energy levels within specific frequency bands. Specifically, increased power in the beta band (approximately 13–30 Hz) has been associated with bradykinesia and hypokinetic states, whereas increased gamma band activity (around 55–75 Hz) is linked to dyskinesia and more active motor states[39 40]. Additionally, the gamma band can entrain to the half harmonic of the stimulation frequency (for example, 65 Hz when using 130 Hz stimulation), offering a potentially stable and reliable reactive biomarker[41 42]. As we have described, multiple morphologies of gamma band have been attributed to PD, including but not limited to broadband, finely tuned, and entrained biomarkers. What each of these means mechanistically and how to best apply them within a classifier is still an active area of research.

While reactive biomarkers are widely studied, predictive biomarkers offer untapped potential. For example, circadian fluctuations in neural signals are highly predictable and can be leveraged for anticipatory adjustments in stimulation. This mirrors strategies used in epilepsy treatment[43] (e.g., vagus nerve stimulation (VNS), albeit for avoiding side effects of sleep apnoea at night[44]), and offers a parallel path for Parkinson's care. By incorporating a real-time clock into the DBS system, it becomes possible to synchronise stimulation patterns with the patient's chronotype and expected symptom patterns throughout the day[45]. In doing so, the system can deliver tailored therapy that proactively addresses symptom exacerbations before they manifest. Similarly, adaptive control during sleep could be modulated to account for overlapping signal features such as sleep spindles that may otherwise confound beta-based biomarkers.

*Risk Mitigation Strategies for a Parkinson's PCLC*

Developing a robust PCLC demands careful attention to risk mitigation strategies. Neural signals used for control are exceedingly small, typically in the range of 1 microvolt root mean square (RMS). This makes them vulnerable to contamination by cardiac artifacts and stimulation interference, both of which can compromise signal integrity. Cardiac interference often overlaps with the beta frequency band, while stimulation artifacts can saturate amplifiers and lead to a failure of the control algorithm. In these scenarios, the adaptive system may default to continuous stimulation, effectively reverting to the established open-loop mode of therapy.

Several strategies are employed to mitigate these risks. Proximity of the implant location to the heart plays a crucial role, and choosing a location further away, preferably on the right side of the body reduces susceptibility to cardiac artifacts[34]. Similarly, careful design of the signal acquisition chain and interleaving of stimulation and sensing functions help prevent amplifier saturation. In alignment with 60601-1-10, additional safeguards include the implementation of safe bounds for stimulation amplitude and the ability to disengage the adaptive algorithm if it fails to operate correctly. These measures ensure that, even in the presence of faults, the system remains within a clinically acceptable and safe operating regime.

As was noted in Figure 2F, increasing stimulation amplitude does not always lead to a decrease in symptoms, nor is the magnitude of dose response constant across stimulation amplitudes. This is possible because other characteristics of stimulation, such as pulse width, also influence physiological response to stimulation[46] (e.g., fibre recruitment depends on charge duration and fibre diameter). Conversely, more obvious non-monotonic behaviour is seen in Parkinson's during hippocampal stim – where if you go a little too high, you go from suppression to an after-discharge and worsened symptoms[47]. Users can mitigate the risk of operating within stimulation doses that exacerbate symptoms by using post-implant symptom titration to identify a patient's dose limits: the dose above which symptoms decrease and the dose above which side effects are observed. The region of operation for the dose curve should be somewhere between these limits.

*Practical Implementation and Performance*

The journey from concept to clinical application has involved a number of key studies. The initial demonstration of adaptive DBS (aDBS) employed a single-threshold "bang-bang" controller that increased or decreased stimulation based on a real-time reading of beta activity[48 49]. This simple reactive approach proved effective in early trials and laid the foundation for more sophisticated control schemes. Subsequent models introduced dual-threshold algorithms, which incorporate upper and lower bounds for neural activity. This dual threshold, "homeostatic" approach reduces unnecessary switching and provides a more homeostatic mode of regulation. Other strategies have explored the use of entrained gamma signals as alternative or complementary biomarkers. These signals, tied to a fixed frequency relative to the stimulation frequency, are less susceptible to cardiac noise and offer a consistent reference point for control. Proportional control methods have also emerged, wherein stimulation levels are scaled in real time according to the magnitude of the beta signal, allowing for more nuanced modulation of therapy. Newronika has an ongoing, double-blind trial with crossover (NCT04681534) directly comparing Parkinson's outcomes with aDBS to those with continuous DBS (cDBS), but results are not yet available.

A recent pivotal trial[50-52] directly compared single-threshold and dual-threshold adaptive controllers against standard continuous DBS. The study achieved its primary outcome, demonstrating that both aDBS approaches maintained similar levels of therapeutic "On" time without troublesome dyskinesia. In terms of raw benefit, the dual-threshold controller yielded a statistically significant and clinically meaningful improvement of +1.3 hours in daily "On" time and -1.6 hours in off time compared to standard DBS[53]. When participants were able to select their preferred method of the options tested, improvements were +1.4 hours in daily "On" time and -1.7 hours in off time compared to standard DBS[54]. Importantly, adverse events were comparable across all treatment modalities, and the majority of participants elected to continue with adaptive stimulation during long-term follow-up.

*What's Next?*

Although the performance gains were moderate, with reduction in total electrical energy delivered (TEED) by aDBS compared to that of cDBS was 15% for the single-threshold approach and 13% for the dual threshold approach, neither of which was statistically significant[53], these results mark a significant step forward in the clinical translation of PCLC systems for PD. With regulatory approvals now in place, future work will focus on refining control algorithms and integrating additional signal features. A particularly promising direction involves the implementation of predictive, time-based control algorithms that differentiate between daytime and nighttime operation. This not only has the potential to reduce side effects during sleep but may also contribute to long-term improvements in sleep architecture, which could have disease-modifying implications.

In conclusion, applying PCLC principles to DBS therapy for PD represents a promising evolution in care. By responding in real time to the patient's neurophysiological state and incorporating safeguards aligned with international safety standards, such systems offer the potential for more personalised, effective, and safer therapy.

**Use Case: Evoked Compound Action Potentials in Spinal Cord Stimulation for Pain**

*Motivation and Mental Model: Stabilizing Dose with Regulation of a Biomarker*

An ECAP is a biopotential originating from a number of activated fibres; this biopotential is elicited by a suprathreshold stimulating pulse. The dorsal spinal ECAP is characterized by a triphasic biopotential resulting from the synchronous activation of dorsal column Aβ fibres in response to an electrical stimulus. The relative size of the ECAP (rather than absolute) is considered to reflect the number of activated fibres; all other factors being constant, as the percent of axons activated increases from zero to 100%, the measured ECAP response will increase monotonically from 0 mV to a relative maximum.

ECAPs may be used to measure the number of activated fibres from (each) SCS stimulus pulse or to measure the changing state of the nervous system (spinal cord) as influenced by the SCS therapy – in the latter case test ECAP stimuli may be interleaved with the therapeutic SCS stimuli [55 56]. These uses may not be independent. The presence (or absence) of an ECAP is not considered a clinical biomarker of pain. The presence (or absence) of an ECAP is not considered necessary or sufficient conditions for pain relief. A stimulation artifact is not an ECAP. A closed-loop SCS system using ECAPs is any system where the stimulation output is adjusted automatically based on measured ECAPs – any such system is PCLC. One such method is to identify a target ECAP level and then adjust stimulation amplitude to maintain the ECAPs at or near the target level (Figure 7).

*Biomarker for Automated Control: ECAP*

Both the generation (neuronal activation) and detection of ECAPs depend on a range of factors, including stimulation dose (e.g. stimulation configuration, pulse width, current), measurement methods (e.g. recording configuration, signal processing, noise), and subject condition (e.g. posture). Some of these factors, such as postural dependence of the distance between the SCS electrodes and the spinal cord, may be leveraged. For example, a change in posture might result in a change in ECAP amplitude which is then used to adjust the stimulation amplitude. ECAP can be leveraged as a Type 2 reactive biomarker (reflecting engagement of mechanisms of therapy action) of spinal cord activation to deliver closed-loop therapy, and our generic schematic for an automated loop can be applied to chronic pain management (Figure 7). A PCLC may then be employed to maintain the amplitude of the ECAP at a target setpoint, by varying the stimulus charge on the next output pulse (Figure 8).

This ECAP biomarker serves as a good example for pragmatic design of a PCLC where physiological interfaces are often variable and non-linear. As a Type 2 biomarker, care must be applied when considering ECAP-controlled PCLC SCS for improved dose consistency. In particular, the characteristics of the sensed ECAP can be influenced by the relative position of the stimulating *and* recording electrode motion with respect to the spinal cord (Figure 8). An increase in spacing between an SCS lead and the cord, for instance, results in a smaller ECAP amplitude for two reasons; when a fixed charge is output from a stimulation electrode, less stimulation energy is coupled from the stimulation electrode to the spinal cord (a smaller number of fibers are activated), and a larger component of the evoked biopotential is shunted to the CSF at the recording electrodes (a smaller sensed ECAP amplitude). These factors manifest as differences in the offset and gain of the dose response curves (Figure 2) for different postures (e.g. see Figure 7 in [11] and Fig. 8 in [57]). Together, these factors mean that a constant ECAP amplitude does not necessarily imply a constant degree of neural activation across multiple postures (e.g. a higher degree of activation is required to achieve a target ECAP amplitude for larger distances between the recording electrodes and the spinal cord)[11 57]. From an intuitive perspective, the change in distance between the spinal cord and the recording electrode introduces an “offset” and “gain error” per Figure 2B. The designer must consider how this source of error might impact the design of their feedback approach. While not negligible, in this example of ECAP-controlled SCS, the relative impact of variations in offset and gain are lower than the inherent variation in stimulation-induced neural activation noted with manual-loop SCS, which is how the control loop can still yield a net benefit in spite of the measurement error (Single et al., 2023). In the parlance of control theory, the ECAP sensor is not strictly linear time invariant (LTI), but in practice, most physiological processes and real-world systems can only be approximated as LTI inside a specific operating window. A common design method is to “linearize” the model for the system around a specific operating point as an approximation (e.g. “the region of interest,” Figure 2C). Since non-LTI feedback systems can underperform, e.g. with a response that oscillates about the control setpoint or otherwise results in an undesired output, the validity of the linear approximation over the operational range of the system should be justified for the control scheme used. Recognizing the limitations of the feedback control variable, such as those described above, is an important consideration for all manner of robust PCLC design and use. Designs using these biomarkers often require a degree of pragmatism, but the limitations of the required assumptions (e.g. linearization about an operating point) and suitable risk mitigations for extending outside the “linear range” must be considered carefully in the context of use.

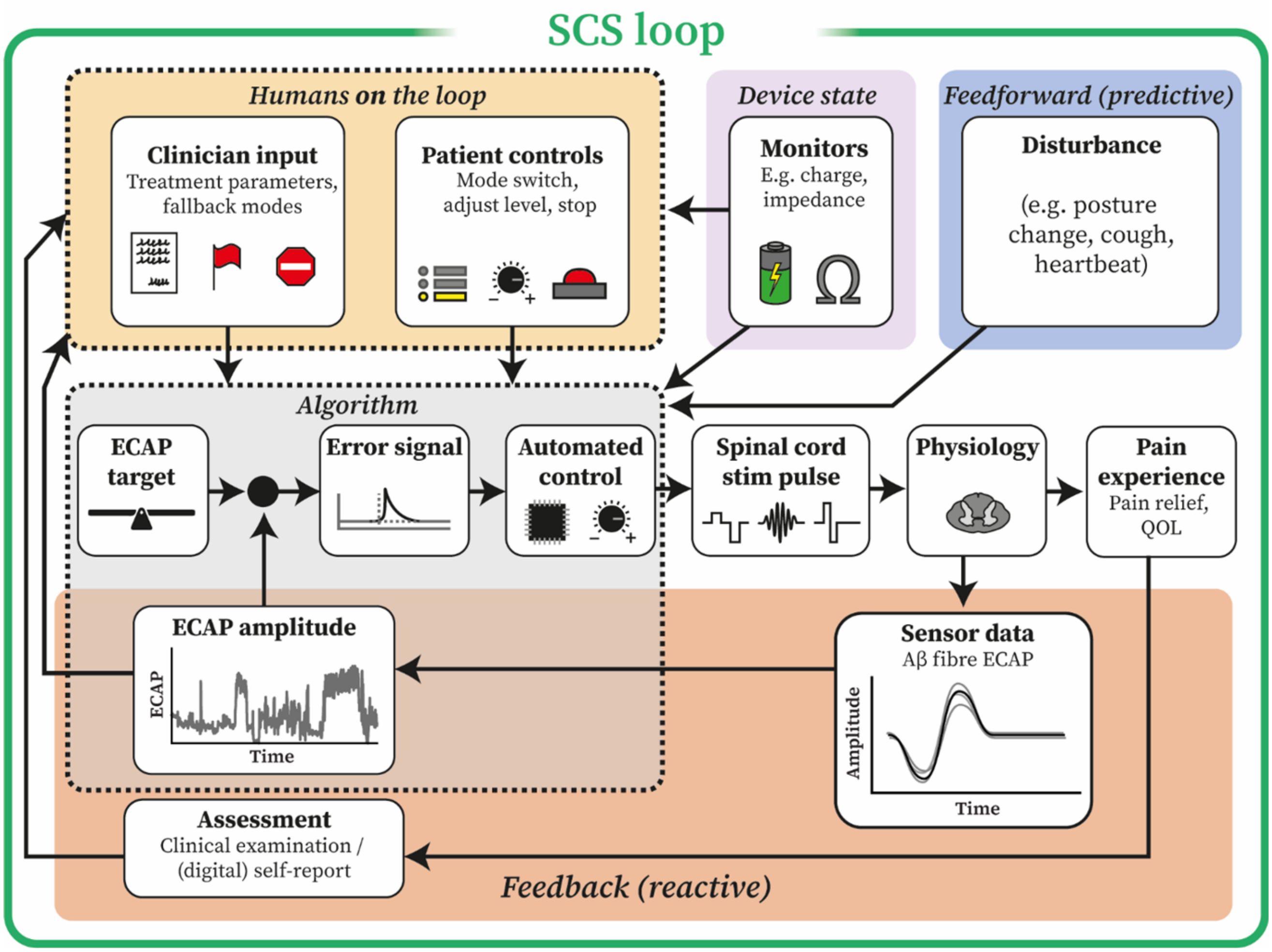

**Figure 7:** Block diagram of a general PCLC system, adapted to the specific examples of the variables used in this case study of SCS for chronic pain treatment. The ECAP is used as a Type 2 reactive biomarker, and changes in posture, among other disturbances) are leveraged as predictive biomarkers.

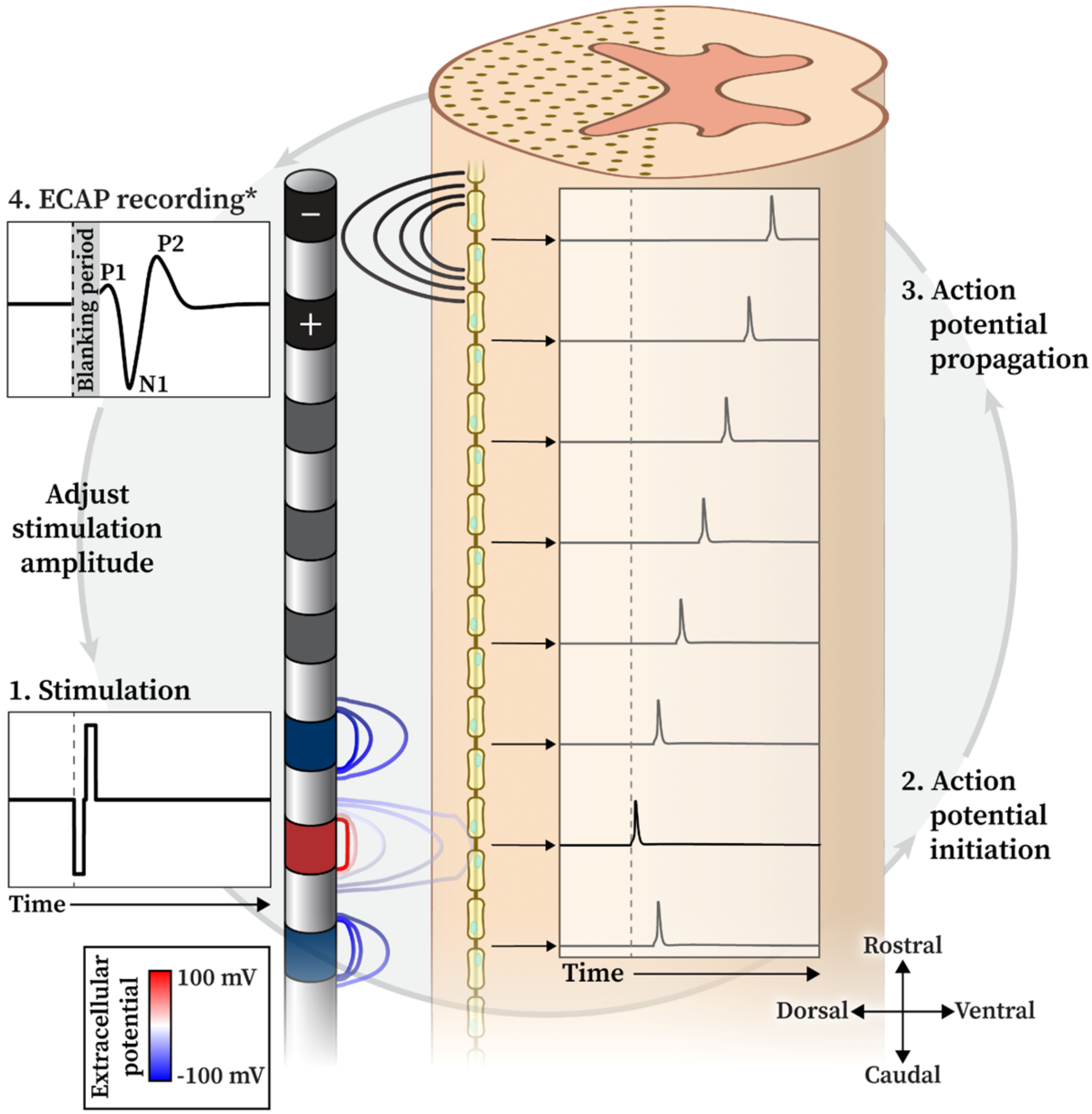

**Figure 8:** PCLC approach to SCS for chronic pain treatment. Depiction of the spinal cord and epidurally placed electrodes, indicating (1) the applied stimulation pulse, (2) action potential generation within the dorsal spinal cord (only a single axon is shown for clarity), (3) action potential propagation along the spinal cord, and (4) bipolar ECAP recording for automatically adjusting the next stimulus pulse amplitude. ∗Importantly, the ECAP represents the summation of action potentials from all active fibres passing by the recording electrodes, rather than the single fibre shown here. Various stimulation configurations (e.g., bipole, guarded cathode) and recording (e.g., adjacent or spaced electrodes) configurations can be utilised provided there is sufficient distance between them to allow for clear separation between the stimulation artifact and neural response. (Reproduced from [11])

*Risk Mitigation Strategies for a Pain PCLC*

While the clinician and patient work together during clinical consultation to set appropriate stimulation limits, this system is able to run freely without continuous clinician involvement. If the patient's experience of their pain does not meet expectations or worsens, then another visit to their clinician can be initiated to adjust parameters further. Any changes in the patient model over time are important to consider; in this case study, data have been published to confirm the long-term stability of the ECAP characteristics and the efficacy of the therapy designed to maintain the ECAP at a target level[58]. Stability of the ECAP at a target level does not imply, however, that the patient is receiving a consistent level of neural activation for the reasons given above, and the approximations for the biomarker use should be factored into the design. Special consideration should also be taken to make sure side effects are minimised with the given parameter sets as well. Specifications should relate to the dose / response relationship of the effect to be optimised or the side effect to be minimised. For example, with a closed-loop system during a cough, the stimulation may exceed the patient's comfort level for continuous stimulation, but if the loop responds quickly enough, the event is not uncomfortable.

An important consideration for the design of a PCLC for SCS is where to implement monitoring of trust checks, or periodic checking of logs that no failure modes or algorithm exit criteria have been encountered and that the system is still operating within assumptions of the mental model. Failure of trust checks should result in an alert and usually should trigger a fallback mode. Regarding the sensor, one must consider sources of noise that disrupt the integrity of the ECAP, such as stimulation artefact, amplifier saturation limits, amplifier noise, broken conductors, or external interference such as airport scanners. A PCLC SCS system should carefully consider how to report or respond to physiologically impossible measurements, such as a negative ECAP amplitude, for instance. If such an event does occur, the PCLC designer should consider what actions to take or not take on them. Regarding the automated control actuator, one must consider the effects of current and voltage compliance limits, resolution (step size), timing resolution and accuracy of the output waveform, tissue encapsulation, and conductor integrity. Additionally, care must be taken that, for example, stimulus artefacts, electrical noise in the ECAP recording system, and environmental noise (such as theft detectors and induction cooktops) do not adversely affect the ability of the loop to maintain the specified level of neural activation.

When trust checks fail, the designer must consider the appropriate strategy to deploy. If external noise is detected that might compromise the ECAP amplitude estimation, the system may revert to manual-loop (fixed output stimulation) mode. In this case, the choice of current setting must be made, including factors such as utilising a last known good state, maintaining therapy, and usability considerations such as alerting the user to the fact that the device is now in manual mode. Design choices also must be made to decide entrance criteria (e.g. duration for which trust checks pass) and appropriate behaviour upon re-entering the automated mode (e.g. which ECAP target setting to maintain).

Usability and user understanding of the operating state of the PCLC should be factored into the design of user interfaces. In this SCS example, the user (usually the clinician) is provided with a real-time trace of the measured signals from the patient (the raw ECAP trace), the estimated ECAP amplitude, the target setpoint, and the stimulation current. This approach allows the user to validate, through real-time visual confirmation, that the loop is sufficiently responsive to postural changes (e.g. the ECAP amplitude time trace doesn't take multiple seconds to recover to the target setpoint), and to ensure the loop is not configured to overshoot the target setpoint (e.g. the ECAP amplitude time trace is oscillating about the target). In this way the user can build a mental model of the relationship between patient movement and PCLC configuration while actively minimising risk.

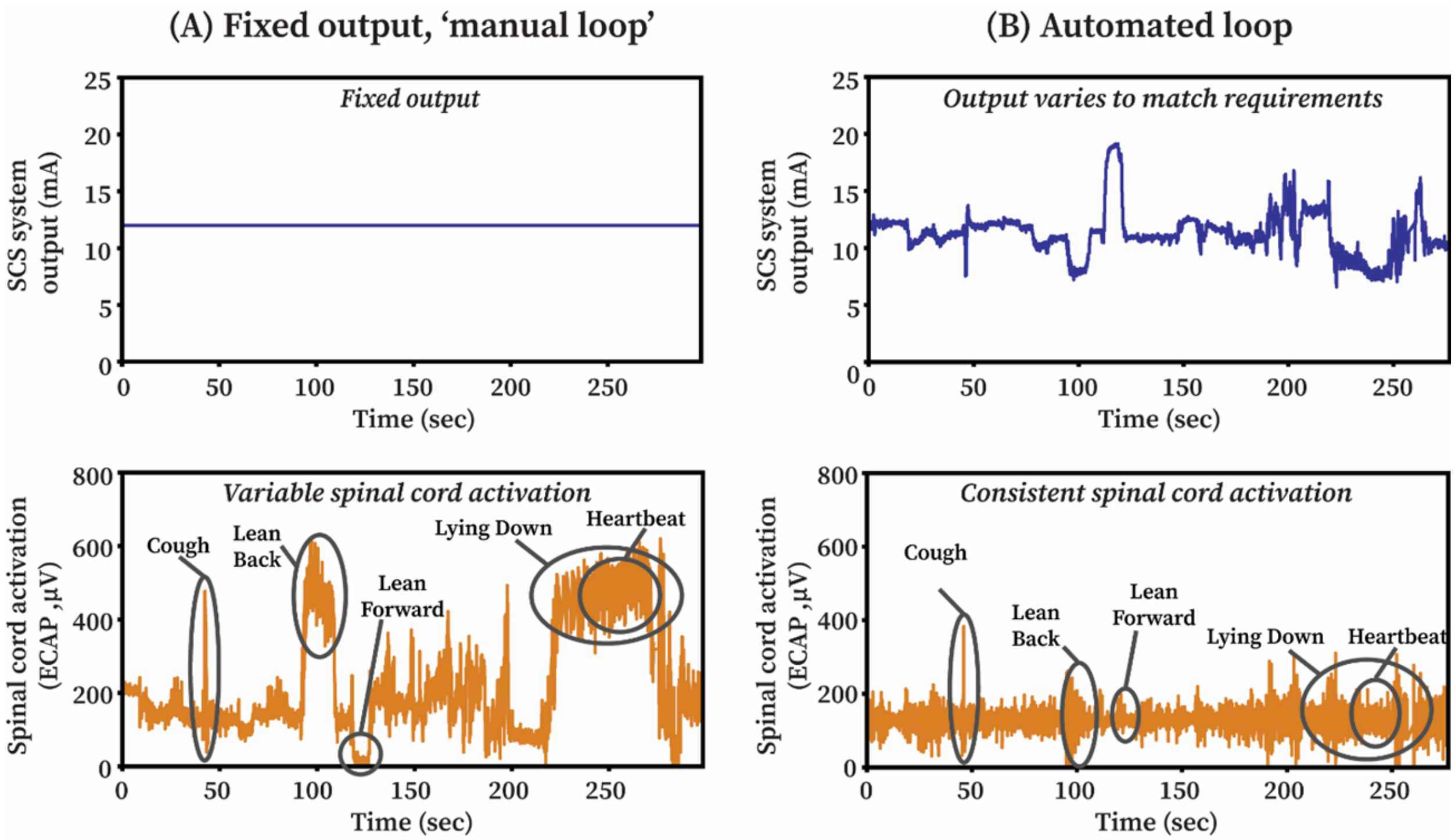

**Figure 9:** Fixed-output, manual-loop SCS (left) versus automated-loop SCS (right). Patient's involuntary physiological and voluntary movements change the separation between the electrode and the spinal cord. Left-hand plots: Fixed-output, manual-loop SCS delivers a fixed output of stimulation current, which leads to a variable level of dorsal column fibre activation during typical daily activities. Right-hand plots: In automated-loop SCS, the stimulation current is automatically adjusted in real time using feedback of the measured neurophysiological response to stimulation to maintain the target response. (Reproduced from [59].)

*Practical Implementation and Performance*

Responsiveness of the PCLC to changes in patient state or ECAP amplitude target setpoint is an important factor in the design of this approach toward PCLC SCS. In SCS, daily activities result in significant variability in the electrode-spinal cord distance. This is illustrated in the "Fixed-output, manual loop" plots of Figure 9, where a fixed output current results in large deviations in dorsal column fibre activation. When a PCLC is deployed to maintain a target ECAP amplitude, the stimulation current must change significantly to cater for the postural changes (see "automated-loop" plots of Figure 9). Standard control system metrics such as response time, settling time, overshoot, and steady-state deviation should be specified and verified; but with a caveat that some of these control system metrics, e.g. "settling time," might assume the system is LTI, and the approximations for this simplified model might warrant alternative metrics depending on the context of use. It is recommended that these parameters are described in terms relevant to the patient; e.g., if our goal is to ensure that the PCLC can effectively respond to the sudden change in electrode-spinal cord distance generated by a cough, then the designer should characterise the expected rate of state change of a cough across a representative population of patients, and design the system to respond in a timeframe that neutralises the impact of this change, or if this data is not available, consider the worst-case situation.

*What's Next?*

As with many PCLC systems, the physiologic variable in this case study (the ECAP) is not a direct biomarker of the disease state being treated (chronic pain), but rather a Type 2, reactive biomarker (a proxy for the number of dorsal columns fibres activated by the stimulus pulse) to assist in SCS during some circumstances. While the ability to record ECAPs is well established[60], lessons continue to be learned through patients implanted with either of the two commercial devices currently leveraging ECAPs in SCS-based pain management[3 58 61 62]. As experience widens, several areas of enhancement are possible: improved sensing to better discriminate signal from noise; compensating for movement of the measurement electrodes; improved supervisory functions to ensure therapy fallback modes are used more rarely; and, a more detailed understanding of the ECAP (dose) to pain relief (response) relationship across patient aetiologies, anatomical variation and inclusive of relevant demographic information. Stimulation adjustments may also be made on a pulse-by-pulse basis when needed as part of the minor loop fine tuning a system with a major loop controlled by one of the emerging "markers of pain"[20]. Recognizing the biophysical basis of the spinal ECAP, ECAP-adjusted PCLC technology has recently been introduced as a tool to help mitigate the number of activated fibres variability intrinsic to all manner of SCS. While potentially relevant to many neuromodulation modalities—such as sacral neuromodulation[63 64]—this PCLC technology is particularly useful with SCS owing to the mobility of the spinal cord and associated changes in the number of activated fibres with activity and postural shifts[11]. Amplitude changes in the ECAP resulting from spinal cord motion may be used as a control signal to inform stimulation adjustments that result in a more consistent number of activated fibres[56 65] and therefore a more predictable response to therapy.

As a field, is it also important to keep searching for a more direct correlate of pain relief, in particular a reactive Type 1 biomarker that directly correlates with the patient's subjective feeling of pain. The bi-directional systems that enable DBS for Parkinson's are being applied to explore these biomarkers and their applications in pilot research studies[66], which more directly close the feedback loop on the clinical symptom of interest.

### Future Use Case: Type 3 Biomarkers

As Type 3 biomarkers measure only energy deposition in the tissue (and not physiological) response they may meet the PCLC requirement "for adjusts or maintains a physiologic variable(s) through delivery or removal of energy (e.g., electric)" but not "using feedback from a *physiologic*-measuring sensor". None the less, a clear and complete rubric of markers used to regulate closed-loop neuromodulation need to classify Type 3. There is emerging research into the relevance of Type 2 markers for therapy. For example, brain impedance (as opposed to simply electrode impedance) for different indications, particularly in epilepsy[67], might prove useful for assessing brain state and adjusting therapy. Impedance metrics are especially useful in concert with well-defined stimulation circuits, and the variance over time and in response to stimulation are potentially useful measures to confirm stimulation was delivered to the target tissue[68].

## Advanced Topics in PCLCs for Neuromodulation

### Training Embedded Algorithms

There is continued interest in algorithms to enable adaptive neurostimulation therapy. The development of automated-loop algorithms requires consideration of the embedded nature of the neurostimulation systems, where onboard computational and memory resources are limited. First, training algorithms requires collection of data to accurately estimate the relationship between biomarkers and symptom severity (e.g., beta band and tremor or bradykinesia in PD) or different neural states (e.g., sleep/wake). Due to the dynamic nature of the nervous system, where physiology varies according to diurnal and longer duration cycles, or the potentially infrequent occurrence of symptoms for different disorders (e.g., seizures in epilepsy), sufficient longitudinal data must be gathered to estimate biomarkers accurately and to characterise their behaviour overtime for good algorithm performance. Key items necessary for training supervised learning algorithms are labelled datasets and known ground truths, both of which may be harder to obtain for some indications than others (consider a binary, objective seizure/non-seizure state label for an epilepsy indication versus a gradient of subjective severities for a pain indication). Continuous data streaming is not desirable for primary cell devices, as this leads to quicker depletion of the device battery and subsequent need for battery replacement surgeries to avoid therapy loss. Rechargeable devices mitigate this concern; however, their battery limits how long they can continuously stream without recharge. Neurostimulation devices may include limited onboard memory to capture data for later download, however use of this requires careful consideration of what data should be captured for later download from the device. As such, there is a need to efficiently capture data so that it can be used for algorithm development. This may be achieved through intermittent download of data recorded to the device embedded memory (for later labelling by trained clinicians) or through external measurement systems and machine learning algorithms for symptom and/or neural state detection to identify periods when data downloading should be triggered. Many PCLCs operate as part of a hierarchical system (Figure 5) that allows for computationally compact algorithms to be embedded on implanted devices but guided by computationally complex algorithms on external devices. In this way, once sufficient data are acquired, algorithm training can be done off the device to identify optimal algorithm parameters to track the biomarker of interest. Said parameters can then be validated on an *in silico* model of the disease state[69] or, alternatively, fed directly back to the implanted device for implementation of the trained algorithm in the closed-loop system. While the intermediary step *in silico* is not mandatory, it is useful from a patient safety point of view to ensure selected parameters do not interact with known elements of the system in a way that defies the mental model of the system.

**Incorporation of Artificial Intelligence**

While devices currently available do not have embedded AI capabilities, it is important to consider how AI fits into existing hierarchically distributed systems and how these will eventually converge. As neuromodulation systems evolve to include sophisticated AI capabilities, it becomes essential to separate the intelligent decision-making components from the safety-critical and infrastructure systems. A modular design paradigm can facilitate this separation by establishing clear boundaries between:

a) *Device Management*: Core functions such as signal acquisition, stimulation delivery, power management, and hardware diagnostics;
b) *Safety and Risk Management*: Regulatory-compliant layers that handle fallback modes, automated control actuator bounds, physiological signal integrity, and alarm systems per ISO 60601-1-10;
c) *Intelligence Layer*: Algorithmic or AI-based systems responsible for interpreting longitudinal data, optimising stimulation strategies, and personalizing care over time.

This layered approach not only improves system resilience and maintainability but also enables incremental upgrades of AI modules without compromising regulatory certification of the base device infrastructure.

In line with the FDA's recommendations for Software as a Medical Device (SaMD), particularly in AI and machine learning (ML) contexts, modularity allows for: 1) *Locked Base Functionality* with traceable version control, meaning the algorithm always generates the same output given the same inputs and does not learn from consecutive uses; 2) *AI model updates* under a pre-specified change control protocol (e.g., FDA's "Predetermined Change Control Plan"); 3) *Explainability-by-design*, where algorithms provide traceable decision paths, allowing clinicians to maintain situational awareness and avoid automation bias; and 4) *Post-market Performance Monitoring* via cloud-aggregated data dashboards.

For safety-critical tasks such as seizure detection or posture-adaptive spinal stimulation, real-time physiological signal interpretation must occur on-device using edge AI deployed between "edge devices" (e.g., sensors, smart phones, and anything else within a specified internet of things). In contrast, non-critical operations, such as retraining models or refining long-term control policies, can be delegated to cloud-based frameworks. This division allows regulatory decoupling, where only validated edge components require full clinical certification, while cloud components may continue evolving under controlled protocols.

As AI components become more deeply integrated into PCLC systems, robust risk mitigation and performance monitoring mechanisms must be embedded throughout the device lifecycle. Regulatory authorities such as the FDA emphasise the importance of model traceability, training data provenance, and ongoing post-market surveillance to ensure that AI driven decisions remain safe, effective, and unbiased over time. Continuous monitoring is particularly critical for models that adapt or are retrained post-deployment. These models must be evaluated against clinically meaningful performance thresholds, and degradation in model accuracy should automatically trigger alerts, human review, or reversion to validated fallback modes. Incorporating confidence scoring, out-of-distribution detection, and real-time audit trails can further reduce the risk of automation bias or silent failure. Such practices align with the regulatory frameworks like FDA's "Good Machine Learning Practice" principles[70] and proposed Total Product Lifecycle (TPLC) approach for AI/ML-based Software as a Medical Device.

Successfully transitioning to modular, AI-enabled PCLC systems requires a systems engineering mindset that embeds safety, usability, and regulatory traceability across all levels of the technology stack. Intelligent systems should remain assistive – not autonomous – unless explicitly validated for autonomous control. Decoupling AI from safety-critical infrastructure enables more rapid innovation, robust oversight, and greater adaptability to clinical demands across diverse patient populations, ultimately resulting in better and more quickly trained embedded algorithms.

**Looking Ahead from Current Challenges in Advanced PCLCs in Diabetes**

The most mature and widely used PCLCs are Automated Insulin Delivery (AID) systems (previously known as Artificial Pancreas Device Systems[71 72]). Over a million people around the world who live with insulin-requiring diabetes depend on these systems to maintain blood glucose in a safe range. AID systems work by sensing blood glucose with a continuous glucose monitor (CGM), deciding how much insulin to deliver with a PCLC algorithm, and acting to deliver the insulin with an insulin pump.

Over the past two decades, AID systems have progressed from in-clinic, laptop-based research setups[73] to fully ambulatory systems integrating components from multiple sponsors. In parallel, the commercial-regulatory ecosystem has transformed, offering lessons – and warnings – that other PCLCs would be wise to heed.

Early AID systems were developed following prospective FDA guidance[72], commercialised by a single sponsor[74], and reviewed by the FDA as Class III[75] medical devices. High costs, intellectual property thickets, and specialised knowledge have grown to the point where it is now too difficult and expensive for a single sponsor to develop all three components: CGM, algorithm, and insulin pump. Today, AID systems are composed from components provided by multiple sponsors. Each component is reviewed as a Class II device with Special Controls[76]:

- **Sense:** 21 CFR 862.1355 Defines the Integrated Continuous Glucose Monitoring System (iCGM), a device that continuously measures glucose and securely transmits data to connected systems[77].
- **Decide:** 21 CFR 862.1356 Defines the Interoperable Automated Glycemic Controller (IAGC), software/hardware that calculates insulin dosing based on glucose and other inputs and sends delivery commands to ACE pumps[78].
- **Act:** 21 CFR 880.5730 Defines the Alternate Controller Enabled Infusion Pump (ACE Pump), insulin pumps designed to interface with iCGMs and iAGCs for automated insulin delivery[79].

iCGM and IAGC devices require clinical validation data gathered via pivotal trial, typically comprising hundreds of subjects for three months. Once a new iCGM or IAGC has been cleared via the 510(k) pathway to be substantially equivalent to the predicate de novo device, they may be integrated into an AID system without additional regulatory review if a clear Predetermined Change Control Plan (PCCP)[80] is in place. ACE pumps only require bench testing and HF validation. All AID systems currently marketed in the United States embed the control algorithm directly in the firmware of the insulin pump.

Some of the challenges faced by the iCGM / IAGC / ACE Pump pathway are described below:

1. The clinical validation, technical integration, and post-market support of IAGC algorithms - some of which are licensed from third parties - are expensive, time consuming, and risky. As a result, it typically takes AID systems 5-15 years to reach widescale commercial availability (if they don’t fail along the way, as most do).
2. Updating an existing IAGC algorithm requires the same clinical validation data and firmware integration as an original algorithm. As a result, there are no “second generation” AID systems yet.
3. The slow pace of commercial-regulatory AID innovation has fostered a #WeAreNotWaiting community[81] of Open Source (OS) Do It Yourself (DIY) user / technologists who have developed their own AID systems beyond the support of sponsors, FDA, or clinicians. The first AID systems to appear in the real world came from this community[82], and they continue to lead commercial AID systems in outcomes[83]. Significant challenges remain in incorporating insights from the DIY community into the commercial-regulatory system.
4. Characterization of safety and efficacy depends on clinical evidence which can only start once hardware is complete and integrated. Then and only then can animal or human testing begin.
5. Despite the enormous pre-market effort to commercialise these systems, most AID systems or their components exhibit problems in the post-market characterised by very high adverse event and product recall rates.
6. Current “interoperable” AID systems are not truly interoperable; they are compatible. Compatibility is achieved through bespoke interfaces: one-off technical and commercial agreements between sponsors, developed at great time and expense, reviewed by FDA. This is very different from “adversarial interoperability / competitive compatibility”[84] which is common in standards-based consumer electronics technologies such as USB and Bluetooth Low Energy, and in other closed loop domains such as oil refinery automation.[85]

7. These systems integrate consumer electronic devices and Internet services and as such can be considered part of the Internet of Medical Things (IoMT)[86]. The rapid pace of Smartphone hardware, operating system, and app development creates challenges for AID systems which depend on these technologies. When new OS updates or cybersecurity vulnerabilities appear, updates and patches must be available quickly.
8. If the original de novo special controls are later found to be missing or incorrect when applied to future 510(k) “substantially equivalent” devices, then it can become difficult or impossible to revise them. For example, consider a new iCGM which exhibits non-physiologic high frequency noise but there was never a special control for high frequency noise in the original iCGM de novo, so it was cleared and integrated in an AID system. If this high frequency noise is now found to negatively affect the IAGC algorithm and user safety, there is no way to remove the device, or to add a one-line-of-code low pass filter, or to add a new special control without incurring significant effort.
9. Since commercial and technical agreements must be reached between sponsors before components can be integrated, it can be time consuming and expensive to gather safety and efficacy evidence. It can be difficult or impossible to evaluate prospective partners.
10. The tremendous cost and risk associated with iCGM and ACE Pump development has consolidated each of these markets into just a few players, who can and do play “market-maker” by restricting access to their components. With few players, intellectual property disputes can jeopardise system access[87]. With few players, a Warning Letter[88], manufacturing problem, or design defect can expose supply chain brittleness.
11. In a multicomponent system with complex feedback interactions, it can be difficult to identify the source of problems. This creates challenges for customer support, for FDA identification of the “Responsible Party”, for the reporting of adverse events, and for the collection of logs and other data which stream from these systems.
12. If a sponsor offers devices in two of the categories – for instance iCGM and IAGC[89] – then the spectre of co-opetition[90] emerges, further complicating commercial and technical relationships.

Various challenges remain with this Class II iCGM / IAGC / ACE Pump pathway, and as a result “It has never been more difficult to launch an insulin pump, CGM, or AID system in the United States. Technical, commercial, and regulatory constraints affect the development and FDA review of AID systems, which must be developed for people from all socioeconomic groups. Cost, risk, and uncertainty of the commercial-regulatory pathway make it very difficult for innovators to participate. They lack access to venture capital and pump or CGM partners. Interoperable AID system development will benefit from an updated, harmonised and transparent commercial-regulatory pathway. Institute of Electrical and Electronics Engineers (IEEE) 11073 and Bluetooth secure plug-and-play interoperability standards will foster true ‘competitive compatibility.’ Alternative evidence generation methods, such as simulation, post-market feedback loops, and FDA acceptance of PRO and RWE studies, would accelerate access to the benefits of AID.”[91] PCLC developers in the neuromodulation space can learn from the experiences of diabetes device developers and avoid pitfalls through early consideration of standards, strategic leverage of modular design, and in silico methods for rapid algorithm validation.

## Innovation and Future Directions

Looking ahead, neuromodulation systems will increasingly integrate multi-layered AI-driven architectures – encompassing implant-level embedded controls, external programmable interfaces, and cloud-based analytics – to enable ongoing refinement and personalization of therapy. This hierarchical control architecture, exemplified by systems like the RNS® System, offers a scalable model for managing complexity and adaptability, allowing devices to evolve continuously through updates driven by patient-specific data and aggregated population insights. As regulatory frameworks evolve alongside these technologies, ensuring validation and safety of dynamic algorithmic updates will become paramount. Ultimately, the fusion of advanced AI with physiological closed-loop neuromodulation presents a transformative potential, setting the stage for precision neurological care capable of real-time, autonomous optimisation tailored uniquely to each patient's evolving clinical landscape. This article may potentially be applicable and useful (on an *ad-hoc* basis) beyond the domain of official PCLC devices, such as for brain computer interfaces and other devices that use various forms of human-in-the-loop control or automated control based on non-physiologic variables. By encouraging researchers and device developers to leverage and apply the PCLC guidance/standards to adjacent device areas, we can hopefully lay the groundwork for future expansion of the FDA Guidance and IEC 60601-1-10 standard to officially include a broader scope of devices.

## Summary Checklist and Conclusions

In this work, we have distilled all neurostimulation systems into either manual or automated loops. We have favoured these terms over “open-loop” and “closed-loop” to instead differentiate based on level of human involvement, or whether the human is more **in** the loop (directly adjusting settings) or **on** the loop (monitoring logs and alerts to ensure proper function). We have provided a generic flowchart (Figure 1) of the components of an automated loop, including opportunities for risk management, and have mapped three examples to this framework. Our intention is that those engineers, physicians, patients, and others using PCLCs utilise this flowchart and the checklist in Figure 10 to ensure good device design and usage. Key points are to know your biomarker and sensor, have a sound mental model of the algorithm and automated control actuator, validate user design and safety features, and provide adequate checks to avoid complacency and skill degradation.

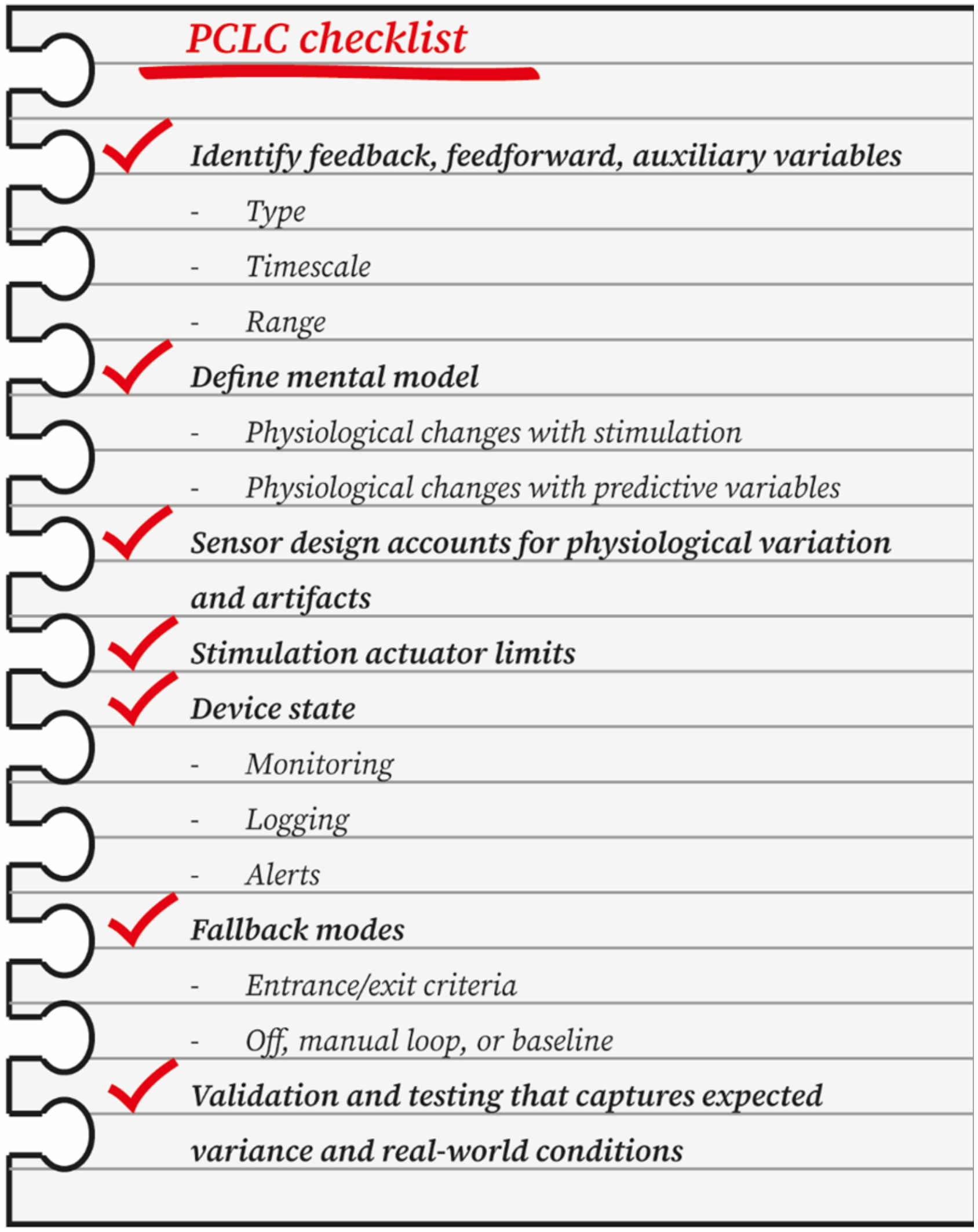

**Figure 10.** Following this checklist will help PCLC developers thoughtfully design safe and effective devices. While this is very high-level, each point has many sub-components discussed at length in the text.

## Acknowledgements

This work is a collaboration across academic and industry partners, sponsored by Institute of Neruomodulation, the non-profit 501c3 research subsidiary of the North American Neuromodulation Society. VSM is funded by the Wellcome Trust. JvR is funded by NIHR i4i. TJD is funded by the Royal Academy of Engineering and the NIHR i4i. We would like to acknowledge Zach McKinney and Dave McMullen for conversations that contributed to the early stages of this work. The authors gratefully acknowledge David Greene and Dr. Thomas K. Tcheng of NeuroPace for their prompt support and assistance in updating and reviewing the description of the NeuroPace RNS® System in this manuscript. Additional support was provided by the NIHR Oxford Health Biomedical Research Centre (NIHR203316). The views expressed are those of the author(s) and not necessarily those of the NIHR or the Department of Health and Social Care.

## Disclosures

VSM was a paid intern at Medtronic, Inc., and a paid consultant at Amber Therapeutics. JvR has received speaker fees from Medtronic, Inc., and is a paid consultant at Amber Therapeutics. DK and PS are employees and shareholders of Saluda Medical. RE was employed by NeuroPace from 2000 to 2016, where she led pivotal advancements in sensing technologies, closed-loop algorithms, and automated detection systems, and contributed to the execution of clinical trials. SFL has received research support from Abbott Neuromodulation, Medtronic PLC, Neuromodulation Specialists, LLC, and Presidio Medical, Inc.; is a shareholder in CereGate, Neuronoff, Inc., and Presidio Medical Inc.,; and is a scientific advisory board member of CereGate and Presidio Medical, Inc..TJD is chief engineer and has shares at Amber Therapeutics, Director at Onward Medical, and a non-executive chairman at Mint Neuro. While this document was prepared with the FDA regulations on Physiological Closed Loop Controllers in mind, no employees of the FDA were involved in the drafting of this document, nor does it serve as any official guidance document. All commentary and advice are based on the authors' experiences and interpretations of the field.